\documentclass[
  twocolumn,
  prl,
  showpacs,
  amsmath,
  amssymb,
  superscriptaddress,
  nofootinbib,
  floatfix
]{revtex4}

\usepackage{mathtools}
\usepackage{bm}
\usepackage{dsfont}
\usepackage{graphicx}
\usepackage{pifont}
\usepackage{textcomp}
\usepackage{gensymb}
\usepackage{accents}
\usepackage{upgreek}
\usepackage{array}
\usepackage{hhline}
\usepackage{tabularx}

\makeatletter
\def\NAT@bibsetnum#1{%
 \setlength{\topsep}{\z@}%
 \NATx@bibsetnum{#1}%
}%
\renewenvironment{thebibliography}[1]{%
 \NAT@thebibliography{#1}%
 \@clubpenalty\clubpenalty
 \let\@TBN@opr\present@bibnote
 \@FMN@list
}{%
 \@endnotesinbib
 \edef\@currentlabel{\arabic{NAT@ctr}}%
 \NAT@endthebibliography
 \global\let\auto@bib\@empty   
}
\newcommand*{\supplementarystart}{%
  \close@column@grid%
  \clearpage%
  \onecolumngrid%
  \setcounter{enumiv}{0} 
  \setcounter{equation}{0} 
  \setcounter{figure}{0} 
  \setcounter{table}{0} 
  \setcounter{page}{1}
  \c@secnumdepth=4
  \renewcommand{\theequation}{s\arabic{equation}} 
  \renewcommand{\bibnumfmt}[1]{[s##1]} 
  \renewcommand{\@onlinecite}{s\citealp} 
  \renewcommand{\cite}[1]{{[}\onlinecite{##1}{]}}
  \renewcommand{\thefigure}{s\arabic{figure}}
  \renewcommand{\thetable}{s\Roman{table}}
  \renewcommand{\thepage}{s\arabic{page}}
}
\makeatother

\newcommand{\be}{\begin{equation}}
\newcommand{\e}{\end{equation}}
\newcommand{\beml}{\begin{subequations}}
\newcommand{\eml}{\end{subequations}}
\newcommand{\beq}{\begin{eqnarray}}
\newcommand{\eq}{\end{eqnarray}}
\newcommand{\ba}{\begin{array}}
\newcommand{\ea}{\end{array}}
\newcommand{\bpm}{\begin{pmatrix}}
\newcommand{\epm}{\end{pmatrix}}
\newcommand{\bc}{\begin{cases}}
\newcommand{\ec}{\end{cases}}
\newcommand{\lt}{\left}
\newcommand{\rt}{\right}

\newcommand{\la}{\langle}
\newcommand{\ra}{\rangle}
\newcommand{\ep}{\varepsilon}

\newcommand{\bb}{\boldsymbol}

\DeclareMathOperator{\tr}{Tr}

\DeclareMathOperator{\im}{Im}
\DeclareMathOperator{\re}{Re}

\DeclareMathOperator{\sign}{sign}

\usepackage{color}

\begin{document}

\title{Anomalous Hall effect in 2D Rashba ferromagnet}

\author{I.\,A.~Ado}
\affiliation{Radboud University, Institute for Molecules and Materials, NL-6525 AJ Nijmegen, The Netherlands}
\author{I.\,A.~Dmitriev}
\affiliation{Max Planck Institute for Solid State Research, Heisenbergstr.\,1, 70569 Stuttgart, Germany}
\affiliation{A.\,F.~Ioffe Physico-Technical Institute, 194021 St.\,Petersburg, Russia}
\author{P.\,M.~Ostrovsky}
\affiliation{Max Planck Institute for Solid State Research, Heisenbergstr.\,1, 70569 Stuttgart, Germany}
\affiliation{L.\,D.~Landau Institute for Theoretical Physics RAS, 119334 Moscow, Russia}
\author{M.~Titov}
\affiliation{Radboud University, Institute for Molecules and Materials, NL-6525 AJ Nijmegen, The Netherlands}

\begin{abstract}
Skew scattering on rare impurity configurations is shown to dominate the anomalous Hall effect in a 2D Rashba ferromagnet. The mechanism originates in
scattering on rare impurity pairs separated by distances of the order of the Fermi wave length. Corresponding theoretical description goes beyond the
conventional non-crossing approximation. The mechanism provides the only contribution to the anomalous Hall conductivity in the most relevant metallic regime and strongly modifies previously obtained results for lower energies in the leading order with respect to impurity strength. 
\end{abstract}

\maketitle

Today we witness a strong revival of interest to spin-orbit induced transport phenomena \cite{Manchon15, Sinova15, Hoffmann13} stimulated in part by increasing role of topology driven issues in condensed matter physics \cite{Hasan2010rev, Qi2011rev}. Experiments with Weyl and Dirac semimetals \cite{Xu15, Xiong15, Kim15} as well as on-going development in the fields of spintronics \cite{Chappert07, Wunderlich10, Jungwirth12, Cheng2013, Joshua2013, Gross2008, Czeschka2011, Liu15}, cold-atoms \cite{Lin11, Cheuk12, Jotzu14}, chiral superconductivity \cite{Xia2006, Xia2008, Kapitulnik2009, Schemm2014}, and magnetisation dynamics \cite{Garello2013, Nagaosa2013, Melnik2014, Fan2014} call for microscopic understanding of the anomalous Hall effect (AHE) \cite{Nagaosa10} that is a key concept uniting these diverse research directions.

The AHE --- a transverse voltage arising in a ferromagnet in response to applied current --- was experimentally discovered as early as in 1881 \cite{Hall81}, but its microscopic origin is still debated \cite{Ado15}. A finite anomalous Hall effect requires the inversion and time-reversal symmetry breaking provided by a combination of the spin-orbit coupling $\alpha$ and magnetization $h$. The celebrated Bychkov-Rashba model \cite{Bychkov84}, see Eq.~(\ref{model}) below, contains just these two nesessary symmetry-breaking terms on top of the usual parabolic kinetic term, and thus is the generic model for theoretical studies of AHE. For this reason, the Bychkov-Rashba model (\ref{model}) and its derivatives are widely used in spintronics for microscopic analysis of the AHE and spin-Hall effects, spin-orbit torques, and other related phenomena \cite{Nagaosa10}. Curiously, extensive studies of this model up to date concluded that the anomalous Hall conductivity $\sigma_{xy}$ vanishes identically above the band 
gap despite being allowed by the symmetry of the Hamiltonian (see Refs.~\onlinecite{Inoue06,Nunner07} and references therein). This remarkable result remains valid even beyond the usual approximation of weak Gaussian impurities \cite{Nunner07}. 

In this Letter, we perform a complete analysis of the AHE in the Rashba ferromagnet in the presense of weak Gaussian impurities. Our treatment incorporates the skew-scattering contribution originating from pairs of close impurities \cite{Ado15} which was not taken into account in previous studies. The explicit calculation shows that full $\sigma_{xy}$ remains \textit{finite} above the gap, see Fig.~\ref{fig:main}. In particular, in the limit of high Fermi energies, $\ep\gg m\alpha^2, |h|$, where $m$ is the effective mass, the resulting conductivity (in units $e^2/2\pi\hbar$) is $\sigma_{xy}=m\alpha^2 h/4\ep^2$, see also Eq.~(\ref{>>}). Thus, in accord with many experiments \cite{Nagaosa10}, the complete theory predicts the scaling $\rho_{xy}\propto \rho_{xx}^2$ of the anomalous Hall resistivity for the metallic ferromagnetic films in the absence of magnetic impurities.

\begin{figure}
\centerline{\includegraphics[width=0.85\columnwidth]{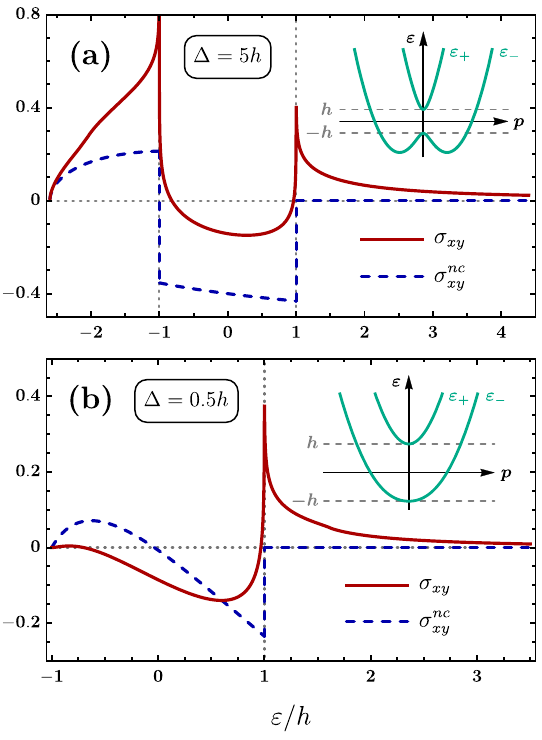}}
\vspace{-8pt}
\caption{Full anomalous Hall conductivity, $\sigma_{xy}=\sigma_{xy}^{\textrm{nc}} + \sigma_{xy}^{\textrm{X}+\rm{\Psi}}$ (solid lines), and the result of the
non-crossing approximation (NCA), $\sigma_{xy}^{\textrm{nc}}$ (dashed lines), in units $e^2/2\pi\hbar$, for model (\ref{model}) as a function of the Fermi energy
$\ep$ for $\Delta\equiv m\alpha^2=5 h$ [panel (a)] and $\Delta=0.5 h$ [panel (b)]. Note that NCA yields $\sigma_{xy}^{\textrm{nc}}=0$ at $\ep>h$ \cite{Inoue06,Nunner07}. 
The corresponding analytical results are summarized in Tables~\ref{table:nc}, \ref{table:result}
and in Eq.~(\ref{>>}). The spectrum of the clean Hamiltonian $H_{\bb{p}}$ is illustrated in insets.}
\vspace{-8pt}
\label{fig:main}
\end{figure}

We consider a 2D system described by the Hamiltonian 
\be
\label{model}
H=H_{\bb{p}}+V,\quad H_{\bb{p}}=p^2/2m+\alpha\, (\bb{\sigma}\times \bb{p})_z  +h\,\sigma_z.
\e
Here the vector $\bb{\sigma}=(\sigma_x,\sigma_y)$ is composed of Pauli matrices, $h>0$ is the exchange field, $\alpha$ the Rashba spin-orbit coupling constant,
and $\hbar=1$. The scalar potential $V=V(\bb{r})$ describes a weak gaussian white-noise disorder with zero average $\la V\ra=0$ characterised by the pair
correlator $\la V(\bb{r}) V(\bb{r}')\ra =  (m\tau)^{-1} \delta(\bb{r}-\bb{r}')$. In this model AHE arises as a result of 
time-reversal symmetry breaking by magnetisation $h$ which affects the electron motion due to spin-orbit coupling
$\alpha$. The resulting anomalous Hall conductivity is an odd function of $h$ and an even function of $\alpha$ due to the symmetry $\sigma_z
H(\alpha)\sigma_z=H(-\alpha)$. 

Our main result for the anomalous Hall conductivity $\sigma_{xy}$ (see Tables~\ref{table:nc},~\ref{table:result} below) is shown by solid lines in
Fig.~\ref{fig:main} for two representative choices of parameters. The dashed lines ($\sigma_{xy}^{\textrm{nc}}$) demonstrate the result  obtained previously
within the non-crossing approximation (NCA) (see Refs.~\onlinecite{Inoue06,Nunner07} and references therein). Recently it was argued  that the NCA misses out an important contribution to $\sigma_{xy}$ which is an inherent part of skew scattering on pairs of impurities \cite{Ado15}. Technically, the missing contribution is represented by the $X$ and $\rm{\Psi}$ diagrams with crossing impurity lines, shown in Fig.~\ref{fig:diag}. 

Parameters in Fig.~\ref{fig:main}a and b correspond to two possible band structures of the clean Hamiltonian $H_{\bb{p}}$, illustrated in the insets of
Fig.~\ref{fig:main}. Eigenvalues of the clean Hamiltonian $H_{\bb{p}}$ correspond to the two spectral branches $\ep_{\pm}(\bb{p})
=p^2/2m\pm\sqrt{\alpha^2p^2+h^2}$. It is therefore convenient to characterise the strength of spin-orbit interaction by the energy scale $\Delta=m\alpha^2$. For $\Delta>h$ one distinguishes three bands with $\ep>h$, $|\ep|<h$, and $\epsilon_{\textrm{min}} < \ep <-h$, where $\epsilon_{\textrm{min}}=
-(h^2+\Delta^2)/2\Delta$, see the inset to Fig.~\ref{fig:main}a. For $\Delta \leq h$ there are two bands (for $\ep>h$ and $|\ep|<h$) and no states below $-h$,
as in Fig.~\ref{fig:main}b. The impurity-crossing mechanism completely determines the AHE in the most relevant regime $\ep>h$.

We calculate the disorder-averaged Hall conductivity as a sum of two contributions $\sigma_{xy}=\sigma_{xy}^\textrm{I}+\sigma_{xy}^\textrm{II}$ using the
Kubo-St\v{r}eda formula \cite{Streda}. At zero temperature these contributions are given by
\beml
\label{Kubo}
\begin{align}
\label{I}
\sigma_{xy}^\textrm{I} &= \frac{1}{2\pi} \tr \la j_x \mathcal{G}^R j_y \mathcal{G}^A\ra,\\
\label{II}
\sigma_{xy}^{\textrm{II}} &= \frac{e}{4\pi i}\tr \la (xj_y-yj_x)\lt(\mathcal{G}^R -\mathcal{G}^A\rt)\ra,
\end{align}
\eml
where $e$ is an electron charge, $\bb{j}=e(\bb{p}/m+\alpha \hat{\bb{z}}\times \bb{\sigma})$, traces include convolution in real or momentum space, angular
brackets stand for the averaging over disorder and $\mathcal{G}^R$, $\mathcal{G}^A$ are exact retarded and advanced Green's functions corresponding to the Hamiltonian $H$ in Eq.~(\ref{model}). 

The quantity $\sigma_{xy}^{\textrm{II}}$ is determined by all electron states below Fermi level. This contribution is insensitive to disorder and can be
rewritten as $\sigma_{xy}^\textrm{II}=ec\, dN/dB$, where  $c$ is the speed of light and $dN/dB$ is the derivative of the total electron concentration $N$ with
respect to magnetic field $B$. The result for $\sigma_{xy}^\textrm{II}$ is quoted in Table~\ref{table:nc} together with other NCA contributions 
\cite{Nunner07}. 

\begingroup
\begin{table}
\begin{center}
\tabcolsep=0.45em
\begin{tabular}{lccc}
\hline\hline\\[-6pt]
&
$\ep>h$ &
$|\ep|<h$ &
$\epsilon_\textrm{min}<\ep<-h$\\[5pt]
\hline\\[-7pt]
$\sigma_{xy}^{\textrm{II}}$ & 0 & $\dfrac{h-\lambda}{2\lambda}$ & $\dfrac{h}{\lambda}$\\[10pt]
$\sigma_{xy}^\textrm{int-I}$ & $\dfrac{-h\Delta}{\lambda_-\lambda_+}$ & $\dfrac{-h\Delta}{2\lambda\lambda_-}$ &
$\dfrac{h\Delta^2}{\lambda\lambda_+\lambda_-}$\\[10pt]
$\sigma_{xy}^\textrm{side}$ & $\dfrac{2h\Delta}{\lambda_-\lambda_+}$ & $\dfrac{2h(\lambda_-^2-h^2)}{\lambda_-(\lambda_-^2+3h^2)}$ & $\dfrac{\lambda
h(\lambda^2-h^2-3\Delta^2)}{(h^2+\Delta^2)\lambda_+\lambda_-}$\\[10pt]
$\sigma_{xy}^\textrm{skew-nc}$ & $\displaystyle\dfrac{-h\Delta}{\lambda_-\lambda_+}$ &
$\displaystyle\dfrac{-3h\lambda(\lambda_-^2-h^2)^2}{2\lambda_-(\lambda_-^2+3h^2)^2}$ & $\displaystyle\dfrac{h \lambda^3
(h^2+2\Delta^2-\lambda^2)}{(h^2+\Delta^2)^2\lambda_+^2\lambda_-^2}$\\[10pt]
\hline\\[-7pt]
$\gamma$ & $\dfrac{1}{2\tau}$ & $\dfrac{\lambda+\Delta}{4\lambda\tau}$ & $\dfrac{\Delta}{2\lambda\tau}$\\[10pt]
$\eta$ & $0$ & $\dfrac{h}{4\lambda\tau}$ & $\dfrac{h}{2\lambda\tau}$\\[10pt]
$\tilde{\alpha}/\alpha$ & $0$ & $1-\dfrac{\lambda(\lambda_-^2-h^2)^2}{\Delta(\lambda_-^2+3h^2)}$ & $1-\dfrac{\lambda^2}{h^2+\Delta^2}$\\[10pt]
$\tau_\textrm{tr}/\tau$ & $1$ & $\dfrac{4\lambda^2}{\lambda_-^2+3h^2}$ & $\dfrac{\lambda^2}{h^2+\Delta^2}$\\[10pt]
\hline\hline
\end{tabular}
\end{center}
\caption{\label{table:nc}
Contributions to the AHE conductivity $\sigma_{xy}^\textrm{nc} = \sigma_{xy}^\textrm{II} + \sigma_{xy}^\textrm{int-I} + \sigma_{xy}^\textrm{side} + \sigma_{xy}^\textrm{skew-nc}$ within the noncrossing approximation (in units $e^2/2\pi\hbar$) for three bands $\ep > h$, $|\ep| < h$, and $\epsilon_\textrm{min} < \ep < -h$ (the latter exists only for $\Delta \equiv m\alpha^2 > h$). The parameters $\lambda = \sqrt{h^2 + 2\ep\Delta + \Delta^2}$, and $\lambda_{\pm} = \lambda\mp\Delta$.}
\end{table}
\endgroup

The main focus of our study concerns the analysis of $\sigma_{xy}^\textrm{I}$ that we calculate perturbatively to the leading order in the parameter  $(\ep_0
\tau)^{-1}\ll 1$, where $\tau$ is the mean scattering time on impurities and $\ep_0$ is the energy difference between the Fermi energy $\ep$ and the closest
band edge. The perturbation theory requires calculation of the Green's function in the leading Born approximation $G^R_{\bb{p}}=(\ep-H_{\bb{p}}-\Sigma^R)^{-1}$,
which yields the self-energy $\im \Sigma^R = -\gamma+\eta \sigma_z$ \cite{sup} with $\gamma,~\eta\propto\tau^{-1}$ given in
Table~\ref{table:nc}. The resulting Green's function can be written as
\be
\label{GR}
G^R_{\bb{p}}=\frac{\ep-s+i\gamma+\alpha\sqrt{2m s}\,\sigma_\phi +(h+i\eta)\sigma_z}{(s-s_+-i\gamma_+)(s-s_--i\gamma_-)}
\e
where $s=p^2/2m$ and $\sigma_\phi = \sigma_x \sin \phi-\sigma_y \cos\phi$
 with the angle $\phi$ pointing in the direction of $\bb{p}$. The terms containing
$\gamma^2$ and $\eta^2$ are disregarded and the parameters $\lambda=\sqrt{h^2+2\ep\Delta+\Delta^2}$, $\lambda_{\pm}=\lambda\mp\Delta$, $s_\pm=\ep\mp
\lambda_\pm$, and $\gamma_{\pm}=(\gamma\lambda_\pm\mp\eta h)/\lambda$ are introduced.

Diagrams contributing to $\sigma_{xy}^\textrm{I}$  in the leading order  are depicted in Fig.~\ref{fig:diag}. The ladder diagram in Fig.~\ref{fig:diag}a yields
$\sigma_{xy}^\textrm{I}\propto(\ep_0 \tau)^{0}$ within the NCA. On the other hand it is well established that the ladder diagram (NCA) provides the leading
Drude result for the longitudinal conductivity $\sigma_{xx}\propto (\ep_0 \tau)^{1}$, while diagrams with intersecting impurity lines are parametrically small
$\propto(\ep_0 \tau)^{0}$. Only recently \cite{Ado15} it was realised that the reasoning validating the NCA for $\sigma_{xx}$ is inapplicable to AHE: the
diagrams with crossing impurity lines in Fig.~\ref{fig:diag}b--d produce additional contributions to $\sigma_{xy}\propto(\ep_0 \tau)^{0}$  
of the same order as the ladder diagram in Fig.~\ref{fig:diag}a.

Physical origin of the failure of the NCA can be understood
using general classification of the AHE mechanisms in terms of intrinsic, side-jump, and skew-scattering contributions. Such
separation naturally arises in the eigenbasis of clean Hamiltonian $H_{\bb{p}}$ and helps to develop an intuitive quasiclassical approach to the effect using
the framework of the Boltzmann kinetic equation \cite{Sinitsyn2007,Sinitsyn2008rev}. The current operator $\bb{j}=e(\bb{p}/m+\alpha\, \hat{\bb{z}}\times
\bb{\sigma})$ does not commute with $H_{\bb{p}}$ and thus has off-diagonal elements in the eigenbasis. The intrinsic part $\sigma_{xy}^\textrm{int-I}$ of
$\sigma_{xy}^\textrm{I}$ results from Eq.~(\ref{I}) with clean Green's functions connecting two off-diagonal current vertices.  
Thus, in the absence of disorder $\sigma_{xy}=\sigma_{xy}^\textrm{int}$, where the total intrinsic Hall conductivity 
 $\sigma_{xy}^\textrm{int}=\sigma_{xy}^\textrm{II}+\sigma_{xy}^\textrm{int-I}$  can be traced
down to the topological properties (Berry curvature) of the Hamiltonian \cite{Junngwirth2002}.
Side jump refers to the transverse displacement of an electron being scattered by impurity; the corresponding contribution $\sigma_{xy}^\textrm{side}$ includes one diagonal and one
off-diagonal vertex in Eq.~(\ref{I}). Finally, skew scattering is due to the asymmetry in the disorder scattering cross-section; it corresponds to 
Eq.~(\ref{I}) with two diagonal current operators.

\begin{figure}
\centering
\centerline{\includegraphics[width=0.8\columnwidth]{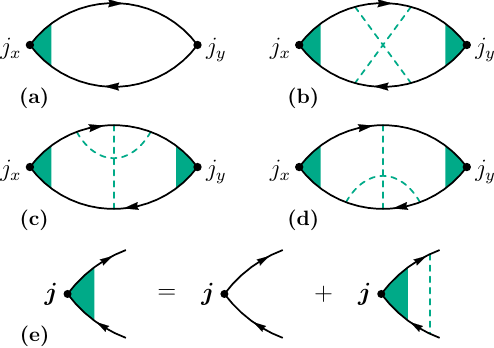}}
\caption{Diagrams for the anomalous Hall conductivity $\sigma_{xy}^\textrm{I}$, see Eq.~(\ref{I}). The non-crossing approximation, diagram (a), yields
$\sigma_{xy}^\textrm{int-I}+\sigma_{xy}^\textrm{side}+\sigma_{xy}^\textrm{skew-nc}$ summarized in Table\ \ref{table:nc} \cite{Nunner07,sup}. The X (b) and
$\rm{\Psi}$ (c,d) diagrams, involving pairs of close impurities, also contribute to the leading order $(\ep_0\tau)^0$, see Table~\ref{table:result}. Vertex
correction (e) involves the sum of ladder diagrams. 
}
\vspace{-0.5cm}
\label{fig:diag}
\end{figure}

The skew-scattering off an individual weak impurity is absent, since its cross-section is symmetric in the Born approximation  \cite{Sinitsyn2007}. Therefore, the skew-scattering in the model (\ref{model}) originates from pairs of impurities at distances of the order of the Fermi wavelength. Such impurity configurations are rare, and they contribute to $\sigma_{xy}$ in the order $(\ep_0 \tau)^{0}$, which is subleading with respect to $\sigma_{xx}\propto(\ep_0 \tau)^{1}$. At the same time, it is the leading order for $\sigma_{xy}$ since both intrinsic and side-jump mechanism also yield  $\sigma_{xy}\propto(\ep_0 \tau)^{0}$. 

The skew-scattering produced by a pair of close impurities is partially included in the NCA diagram in Fig.~\ref{fig:diag}a \cite{Nunner07,sup}. However, proper treatment of the coherent scattering by such close pairs must also include diagrams in Fig.~\ref{fig:diag}b--d missing in the NCA, since at short distances crossing of impurity lines does not produce extra smallness. Thus the X and $\rm{\Psi}$ diagrams represent an inherent part of skew scattering parametrically indistinguishable from the NCA part, Fig.~\ref{fig:diag}a. 

It is important to mention that since impurities are close to each other, the skew scattering on impurity pairs is largely insensitive to temperature, magnetic field, and other decoherence mechanisms.
On the other hand, diagrams with more than two crossing lines (weak localization) are sensitive to decoherence but contain an extra smallness hence will be disregarded.

\begingroup
\begin{table*}
\begin{center}
\tabcolsep=0.5em
\begin{tabular}{ll}
\hline\hline\\[-6pt]
Region & $\sigma_{xy}^{\textrm{X}+\rm{\Psi}}=\sigma_{xy}^{\textrm{X}} +\sigma_{xy}^{\rm{\Psi}}$ (in units $e^2/2\pi\hbar$) \\[5pt]
\hline\\[-7pt]
$\varepsilon>\epsilon_+$ & $\dfrac{h\Delta}{\pi a \lambda ^3}\Big[(2\Delta+\lambda) \sqrt{1+2 a}\left(K_+-E_+\right)+(2\Delta-\lambda) \sqrt{1-2
a}\left(K_--E_-\right)\Big]$\\[10pt]
$h<\varepsilon<\epsilon_+$ & $\dfrac{h\Delta}{\pi a \lambda ^3}\Big[(2\Delta+\lambda) \sqrt{1+2 a}\left(K_+-E_+\right)-(2\Delta-\lambda) \sqrt{2
a-1}E_-'\Big]$\\[10pt]
$|\,\varepsilon\,|<h$ & $\dfrac{4 h \lambda \Delta}{\pi(\lambda_-^2 + 3 h^2)^2} \Bigl[(2\Delta + \lambda) \bigl[ \pi + 2 \sqrt{1 + 2a} (K_+ - 2 E_+) \bigr] - \pi\lambda a (1 + a) \Bigr]$\\
$\epsilon_-<\varepsilon<-h$ & $\dfrac{h \lambda \Delta}{\pi  \left(h^2+\Delta^2\right)^2}\left[\dfrac{2\Delta+\lambda}{a} \sqrt{1+2
a}\left(K_+-E_+\right)+\dfrac{2}{3}(2\Delta-\lambda)\left(\pi+(2-a)\sqrt{2 a-1}K_-'\right)\right]$\\[10pt]
$\epsilon_{min}<\ep<\epsilon_-$ & $\dfrac{h \lambda \Delta}{\pi  \left(h^2+\Delta^2\right)^2}\left[\dfrac{2\Delta+\lambda}{a} \sqrt{1+2
a}\left(K_+-E_+\right)+\dfrac{2}{3}(2\Delta-\lambda)\left(\pi-(2-a)\sqrt{1-2 a}K_-\right)\right]$\\[10pt]
\hline\\[-7pt]
Notations: & $\Delta\equiv m\alpha^2$,\quad\qquad$\lambda=\sqrt{h^2+2\ep\Delta+\Delta^2}$,\quad\qquad$\epsilon_\pm = (9\Delta\pm
5\sqrt{9\Delta^2+16h^2})/16$;\\[10pt]
$a=\dfrac{s_{-}-s_{+}}{s_{-}+s_{+}+2\sqrt{s_{+}s_{-}}},\quad|\ep|>h$; & $a=\dfrac{1}{2}\lt[\sqrt{\dfrac{9s_{-}-s_+}{s_{-}-s_{+}}}-1\rt],\quad|\ep|<h$; \qquad\qquad$s_\pm=\ep+\Delta\mp\lambda$;
\\[10pt]
Elliptic $K_\pm=K(k_\pm)$,\quad $K'_\pm=K(k'_\pm)$,& $E_\pm=E(k_\pm)$,\quad $E'_\pm=E(k'_\pm)$\; with \;$k_\pm^2=a^3(2\pm a)/(2a\pm 1)$ and
$(k'_{\pm})^2=1-k_\pm^2$.\\[5pt]
\hline\hline
\end{tabular}
\end{center}
\caption{\label{table:result}
The part $\sigma_{xy}^{\textrm{X}+\rm{\Psi}}$ of the leading-order anomalous Hall conductivity $\sigma_{xy}$ not captured by the noncrossing approximation. The
analytic results for 5 distinct energy regions in the three energy zones indicated in Table~\ref{table:nc} are shown along with notations in terms of the
parameters of the model (\ref{model}).}
\end{table*}
\endgroup

Calculation of $\sigma_{xy}^\textrm{nc}$ goes along the lines of Ref.~\onlinecite{Nunner07}. Summation of the ladder diagrams in Fig.~\ref{fig:diag}e yields the
dressed current operator. In the leading order, it takes the form $\bb{j}^\textrm{dress}=e(\bb{p}/m+\tilde\alpha\, \hat{\bb{z}}\times \bb{\sigma})$ with the
modified parameter $\tilde\alpha$ given in Table~\ref{table:nc}. This is sufficient for the calculation of the diagonal conductivity $\sigma_{xx}$ while the
calculation of $\sigma_{xy}^\textrm{nc}$ requires the subleading term in $\bb{j}^\textrm{dress}$ that has a different matrix structure $\sim \bb{\sigma}$.
Details of the calculation are given in Supplementary Material \cite{sup}. Separation of the result into individual contributions is summarised in
Table~\ref{table:nc}.

The contributions of X and $\rm{\Psi}$ diagrams in Fig.~\ref{fig:diag}b--d are most easily evaluated \cite{sup} in real space. The Green function (\ref{GR}) is decomposed in the clean limit, $\tau\to \infty$, into terms corresponding to the two spectral branches,
\beml
\label{GG}
\begin{align}
G^R_{\bb{r}} &= G^R_+(\bb{r})+G^R_-(\bb{r}),\\
G_{\pm}^R &=\pm \frac{1}{2\lambda} \lt[\ep+\frac{\nabla^2}{2m} -i\alpha\,\bb{\sigma}\!\times\!\bb{\nabla} +h\sigma_z\rt]g_{\pm}(r).\;
\end{align}
\eml
Here we introduce the functions
\beml
\label{gg}
\begin{align}
g_{-} &= \frac{1}{2}\lt[Y_0(p_-r)-iJ_0(p_-r)\rt],\\
g_{+} &= \bc \frac{1}{2}\lt[Y_0(p_+r)-iJ_0(p_+r)\sign\ep\rt],&|\ep|>h,\\
-\frac{1}{\pi}K_0(|p_+|r),&|\ep|<h, \ec\quad
\end{align}
\eml
and $J_0$, $Y_0$, and $K_0$ stand for the standard Bessel functions. We use the notation $p_\pm = \sqrt{2m s_\pm}$ for the two Fermi momenta.

The diagrams with crossed impurity lines can be represented as 
\beml
\label{XY}
\begin{align}
\sigma_{xy}^\textrm{X} &= \frac{1}{2\pi} \frac{1}{(m\tau)^2} \int d^2\bb{r} \tr\lt[J^x_{\bb{r}} G^R_{-\bb{r}} J^y_{\bb{r}} G^A_{-\bb{r}}\rt],\\
\sigma_{xy}^{\rm{\Psi}} &= \frac{1}{\pi} \frac{1}{(m\tau)^2} \re \int d^2\bb{r} \tr\lt[J^x_{\bb{r}} G^R_{-\bb{r}} G^R_{\bb{r}}J^y_{-\bb{r}}\rt].
\end{align}
\eml
The function $\bb{J}_{\bb{r}} = [ G^A \bb{j}^\textrm{dress} G^R ]_{\bb{r}}$ is given by
\be
\label{J}
 \bb{J}_{\bb{r}}
 = e \tau_\text{tr} \nabla \Bigl[
      G_-^R - G_-^A + \Bigl( G_+^R - G_+^A \Bigr)\mathop{\mathrm{sign}} \ep
    \Bigr],
\e
with the parameter $\tau_\text{tr}$ from Table~\ref{table:nc}. All Green's functions in Eqs.~(\ref{XY},\ref{J}) can now be taken from Eqs.~(\ref{GG},\ref{gg}), i.\,e.\,in the leading order with respect to $\tau$.

All integrations involved in Eqs.~(\ref{XY}) can be performed analytically \cite{sup}. The final result, given in
Table~\ref{table:result}, is written in terms of an auxiliary parameter $0<a<1$, which takes on the value $1$ for $\ep=\pm h$ and is vanishing at
$\ep=\epsilon_\textrm{min}$ and for $\ep\to \infty$. Complete elliptic integrals of the first and second kind arising in Eqs.~(\ref{XY}) have moduli $k_\pm^2$
or $(k'_{\pm})^2=1-k_\pm^2$ also quoted in Table~\ref{table:result}. The value of $k_+$ is real and restricted to $0<k_+<1$. The value of $k_-$ is imaginary for
$0<a<1/2$ and real for $1/2<a<1$. Switching between these two regimes occurs at energies $\epsilon_\pm$ which further divide the spectrum into five different
regions. The values $\ep=\epsilon_\pm$ correspond to the ``nesting'' configuration such that $p_-=3 p_+$. Note that for $h>\Delta$ there are no states below
$-h$ and the last two cases in Table~\ref{table:result} are absent.

The overall result for the anomalous Hall conductivity $\sigma_{xy}=\sigma_{xy}^{\textrm{nc}} + \sigma_{xy}^{\textrm{X}+\rm{\Psi}}$ is shown in
Fig.~\ref{fig:main}. NCA yields jump discontinuities in the Hall conductivity at $\ep=\pm h$. Proper treatment of the skew scattering including X and
$\rm{\Psi}$ contributions introduces logarithmic singularities at the same points. However, the approximations used in our calculation are invalid in the narrow
vicinity of these two energies, when the smallest Fermi momentum is comparable to the inverse mean free path. Therefore, the logarithmic singularities are
artificial.

At large energies $\ep\gg h,\Delta$, the anomalous Hall conductivity $\sigma_{xy}=\sigma_{xy}^{\textrm{X}+\rm{\Psi}}$ can be expanded as (in units $e^2/2\pi\hbar$)
\begin{equation}
\label{>>}
 \sigma_{xy}
  = \frac{h \Delta}{4}\left[\frac{1}{\varepsilon^2} - \frac{\Delta}{\varepsilon^3} + \frac{3 (7\Delta^2 + 8 h^2)}{32 \varepsilon^4} + \ldots
    \right].
\end{equation}
The most striking feature of the X and $\rm{\Psi}$ contributions is the fact that they give rise to non-zero AHE for $\ep > h$ where
$\sigma_{xy}^{\textrm{nc}}=0$.  In Ref.~\onlinecite{Nagaosa10} the term ``skew scattering'' refers exclusively to the skew scattering off strong single
impurities that contributes to $\sigma_{xy}$ in the order $\ep_0\tau$. Single-impurity skew-scattering manifests itself in resistivity $\rho_{xy}\propto
\rho_{xx}$ but is generally absent in the limit of Gaussian disorder. The model (\ref{model}) is, however, special since the NCA conductivity
$\sigma_{xy}^{\textrm{nc}}$ does vanish for $\ep>h$ even beyond Gaussian approximation \cite{Nunner07}. Non-vanishing AHE at $\ep>h$ was reported to appear 
only when the forth-order non-Gaussian disorder correlations are taken into account \cite{Kovalev08,Kovalev09}.

Quite generally the absence of skew-scattering on single impurities manifests itself in the scaling $\rho_{xy}\propto \rho_{xx}^2$ that can be tested
experimentally by varying impurity concentration. We stress that whenever such scaling takes place the skew-scattering on rare impurity pairs have to be taken into account, which necessarily involves the analysis of X and $\rm{\Psi}$ contributions to AHE.

Our results may be of direct relevance for recent experiments with LaAlO$_3$/SrTiO$_3$ interfaces and ferromagnet-platinum bilayers \cite{Cheng2013, Joshua2013, Gross2008, Czeschka2011, Liu15} . A spin-orbit induced valley-Hall effect of similar kind may also be observed in graphene on WS$_2$ \cite{Avsar2014} and in synthetic systems such as ultra-cold Fermi gases \cite{Cheuk12}. A closely related phenomenon, the spin-orbit torque on magnetisation \cite{Manchon2009, Miron2010}, is also strongly affected by skew-scattering on rare impurity configurations and calls for similar analysis.

In conclusion, we reconsidered the anomalous Hall effect in 2D Rashba ferromagnet in the presence of weak impurities. Our diagrammatic approach fully takes into account the skew scattering which requires going beyond the conventional noncrossing approximation. In sharp contrast to previous knowledge \cite{Sinitsyn2007, Sinitsyn2008rev, Nagaosa10}, such complete treatment yields a finite anomalous Hall conductivity in the metallic regime (\ref{>>}). This implies the $\rho_{xy}\propto \rho_{xx}^2$ scaling of the anomalous Hall resistivity for metallic ferromagnetic films in the absence of magnetic impurities.

We are grateful to Gerrit Bauer, Aurelien Manchon, and Jairo Sinova for helpful discussions. The work was supported by the Dutch Science Foundation NWO/FOM 13PR3118 and by the EU Network FP7-PEOPLE-2013-IRSES Grant No 612624 ``InterNoM''. Results from Table~\ref{table:nc} were obtained with support from Russian Science Foundation (Grant No.\ 14-42-00044).

\supplementarystart

\centerline{\bfseries\large ONLINE SUPPLEMENTAL MATERIAL}
\vspace{6pt}
\centerline{\bfseries\large Anomalous Hall effect in 2D Rashba ferromagnet}
\vspace{6pt}
\centerline{I.\,A.~Ado, I.\,A.~Dmitriev, P.\,M.~Ostrovsky, and M.~Titov}
\begin{quote}
In this Supplemental Material we provide technical details that are relevant for the text of the Letter. In particular, we explain the separation of the Hall
conductivity into intrinsic, side jump, and skew scattering contributions, and explicitly calculate the X and $\Psi$ diagrams both in momentum and real space
representation.
\end{quote}

\maketitle

\section{Disorder-averaged Green function}

%

The Green function for the Hamiltonian (\ref{model}) of the main text acquires a self energy $\Sigma^{R,A} = \mp i(\gamma - \eta \sigma_z)$, when averaged with
respect to disorder, and takes the form
\begin{equation}
\label{GRAsup}
 G^{R,A}
  = [\ep - \Sigma^{R,A} - H_{\boldsymbol{p}}]^{-1}
  = \frac{\ep - p^2/2m \pm i\gamma + \alpha (\bm{\sigma} \times \mathbf{p})_z + (h \pm i\eta) \sigma_z}
         {(\ep- p^2/2m \pm i \gamma)^2 - \alpha^2 p^2 - (h \pm i\eta)^2}
  = \frac{\ep - s \pm i\gamma + \alpha \sqrt{2 m s}\, \sigma_\phi + (h \pm i\eta) \sigma_z}{(s - s_+ \mp i\gamma_+)(s - s_- \mp i\gamma_-)}.
\end{equation}
This reproduces Eq.\ (\ref{GR}) of the main text. We express the Green function in terms of the variable $s = p^2/2m$ and momentum direction $\phi$ with
$\sigma_\phi = \sigma_x \sin\phi - \sigma_y \cos\phi$. The denominator is factorised in $s$ with the help of the following notations:
\begin{gather}
\Delta = m\alpha^2, \qquad
\lambda = \sqrt{h^2 + 2 \ep \Delta + \Delta^2}, \qquad
\lambda_\pm = \lambda \mp \Delta, \\
s_\pm = \ep\mp \lambda_\pm, \qquad\qquad
\gamma_\pm = (\gamma \lambda_\pm \mp \eta h)/\lambda,
\end{gather}
which are also used in the main text of the Letter (cf.\ Table \ref{table:result}). We note that the spectral branches are defined such that $s_+ < s_-$ and
$s_- > 0$ while $\mathop{\mathrm{sign}} s_+ = \mathop{\mathrm{sign}}(|\ep|^2 - h^2)$.

To simplify subsequent calculations, we take advantage of dimensionless notations by letting $m = \alpha = \hbar = 1$. Thus the energy variables, such us
$\lambda$ and $h$, are measured in units of the parameter $\Delta$. We also express the Fermi energy through the variable $\lambda$ as $\ep= (\lambda^2 -
h^2 - 1)/2$. With these conventions, all results are expressed in terms of the two parameters $h$ and $\lambda$. In the dimensionless notations, we also have
$\lambda_\pm = \lambda \mp 1$.

\subsubsection*{Reduction of denominators}

We will use separate notations for the numerator and denominator of the Green function in the representation of Eq.\ (\ref{GRAsup}), $N^{R,A}$ and $D^{R,A}$,
respectively. The denominators depend only on the variable $s$ and obey the following useful identities:
\begin{align}
 \frac{1}{D^R} - \frac{1}{D^A} 
 &= \frac{i \pi}{\lambda} \Bigl[
      \mathop{\mathrm{sign}} \gamma_-\, \delta(s - s_-)
      - \mathop{\mathrm{sign}} \gamma_+\, \delta(s - s_+)
    \Bigr] + O(\gamma_\pm), \label{dendiff} \\
 \frac{1}{D^R D^A}
  &= \frac{\pi}{4 \lambda^2} \left[
      \frac{\delta(s - s_-)}{|\gamma_-|} + \frac{\delta(s - s_+)}{|\gamma_+|}
    \right] + O(1). \label{denden}
\end{align}
The numerators are given by $N^{R,A}=\tfrac{1}{2}(\lambda_+\lambda_--h^2)\pm i\gamma+\sqrt{2s}\sigma_\phi+(h\pm i\eta)\sigma_z$.

\subsection{Scattering rate}

The parameters of the self-energy $\gamma$ and $\eta$ are obtained in the Born approximation from the following equation:
\begin{equation}
 \gamma - \eta \sigma_z
  = \frac{i}{2} \bigl[ \Sigma^R - \Sigma^A \bigr]
  = \frac{i}{2 \tau} \int \frac{d^2p}{(2\pi)^2} \bigl[ G^R(\mathbf{p}) - G^A(\mathbf{p}) \bigr].
\end{equation}
After angular integration, the integrand depends on the variable $s$ only. Disregarding $\gamma$ and $\eta$ in the right-hand side and using Eq.\
(\ref{dendiff}), we obtain
\begin{equation}
 \gamma - \eta \sigma_z
  = \frac{1}{4 \lambda \tau} \Bigl[
      (\lambda_- - h \sigma_z) \mathop{\mathrm{sign}} \gamma_- \theta(s_-) + (\lambda_+ + h \sigma_z) \mathop{\mathrm{sign}} \gamma_+ \theta(s_+)
    \Bigr].
\end{equation}
This equation yields the result,
\begin{equation}
\label{Born}
 \gamma
  = \frac{1}{4 \lambda \tau} \begin{cases}
      2\lambda, & \ep> h, \\
      \lambda_-, & |\ep| < h, \\
      2, & \ep< -h,
    \end{cases}
 \qquad
 \eta
  = \frac{h}{4 \lambda \tau} \begin{cases}
      0, &\ep> h, \\
      1, &|\ep| < h, \\
      2, & \ep< -h,
    \end{cases}
 \qquad
 \gamma_\pm
  = \frac{1}{4 \lambda^2 \tau} \begin{cases}
      2\lambda_\pm \lambda, &\ep> h, \\
      \lambda_\pm \lambda_- \mp h^2, &|\ep| < h, \\
      2\lambda_\pm \mp 2 h^2, & \ep< -h.
    \end{cases}
\end{equation}
These scattering rates are listed in Table \ref{table:nc} of the main text. It is useful to remember that $\gamma_- > 0$ and $\mathop{\mathrm{sign}} \gamma_+ =
\mathop{\mathrm{sign}} \ep$.

\subsection{Density of states}

As a byproduct of the above calculation, we obtain the density of states
\begin{equation}
 \rho
  = -\frac{1}{\pi} \mathop{\mathrm{Im}} \mathop{\mathrm{Tr}} G^R
  = \frac{2 \tau \gamma}{\pi}
  = \frac{1}{\pi} \begin{cases}
      1, & \ep> h, \\
      \lambda_-/2\lambda, & |\ep| < h, \\
      1/\lambda, & \ep< -h,
    \end{cases}  
\end{equation}
in all three (or two, if $h < 1$) spectral bands.

\section{Hall conductivity in the non-crossing approximation}


\subsection{Vertex corrections}

Hall conductivity is expressed by the Kubo formula, Eqs.\ (\ref{Kubo}) of the main text. The contribution $\sigma_{xy}^\text{II}$ is insensitive to disorder 
and was computed in Ref.\ \cite{CommonGrave}; the result is given in Table\ \ref{table:nc}. Here we focus on the calculation of $\sigma_{xy}^\text{I}$, Eq.\ 
(\ref{I}), in the non-crossing approximation. It amounts to calculation of the disorder ladder diagrams for vertex corrections, see Fig.\ \ref{fig:diag}e. 

Bare current operator contains kinetic and spin-orbit parts (we use the units with $m = \alpha = 1$):
\begin{equation}
 \mathbf{j}
  = e\, (\mathbf{p} + \hat{\mathbf{z}} \times \bm{\sigma}).
 \label{j}
\end{equation}
We introduce the following four quantities to describe current dressing by disorder:
\begin{align}
 \int \frac{d^2p}{(2\pi)^2}\; G^R \mathbf{p} G^A
  &= A\; \hat{\mathbf{z}} \times \bm{\sigma} + C\; \bm{\sigma}, \\
 \int \frac{d^2p}{(2\pi)^2}\; G^R (\hat{\mathbf{z}} \times \bm{\sigma}) G^A
  &= B\; \hat{\mathbf{z}} \times \bm{\sigma} + D\; \bm{\sigma}.
\end{align}
Upon averaging with respect to momentum directions, the integrands of the above expressions depend on $s$ only,
\begin{align}
A&= \int \frac{ds}{2\pi}\; \frac{2 s (\ep- s)}{D^R(s) D^A(s)}, &
B&= \int \frac{ds}{2\pi}\; \frac{(\ep- s)^2 - h^2}{D^R(s) D^A(s)}, \\
C&= -\int \frac{ds}{2\pi}\; \frac{2 \eta s}{D^R(s) D^A(s)}, &
D&= \int \frac{ds}{2\pi}\; \frac{2 \gamma h - 2\eta (\ep- s)}{D^R(s) D^A(s)}.
\end{align}
Note that the integrals $A$ and $B$ are of the order $O(\tau)$ while $C$ and $D$ are of a subleading order $O(1)$. Using the representation of
Eq.~(\ref{denden}) for the denominators of the Green functions and the Born values of scattering rates Eq.\ (\ref{Born}), we obtain the following results:
\begin{align}
 A&= -\frac{\tau}{2} \begin{cases}
      2, & \ep> h, \\
      \dfrac{\lambda_- (\lambda_-^2 - h^2)}{\lambda_-^2 + h^2}, & |\ep| < h, \\
      \dfrac{\lambda^4 - 4 \lambda^2 h^2 + h^4 - 1}{(\lambda_- + h^2)(\lambda_+ - h^2)}, & \ep< -h,
    \end{cases} &
 B&= \frac{\tau}{2} \begin{cases}
      \dfrac{\lambda^2 - h^2 - 1}{\lambda^2 - 1}, & \ep> h, \\
      \dfrac{\lambda_-^2 - h^2}{\lambda_-^2 + h^2}, & |\ep| < h, \\
      \dfrac{(h^2 - 1)(h^2 - \lambda^2 + 1)}{(\lambda_- + h^2)(\lambda_+ - h^2)}, & \ep< -h,
    \end{cases} \\
 C&= -\frac{h}{8 \lambda} \begin{cases}
      0, & \ep> h, \\
      \dfrac{\lambda_-^2 - h^2}{\lambda_-^2 + h^2}, & |\ep| < h, \\
      \dfrac{2(h^2 - 1)(h^2 - \lambda^2 + 1)}{(\lambda_- + h^2)(\lambda_+ - h^2)}, & \ep< -h,
    \end{cases} &
 D&= \frac{h}{2 \lambda} \begin{cases}
      \dfrac{\lambda}{\lambda^2 - 1}, & \ep> h, \\
      \dfrac{\lambda_-}{\lambda_-^2 + h^2}, & |\ep| < h, \\
      \dfrac{\lambda^2 - 2 h^2 - 2}{(\lambda_- + h^2)(\lambda_+ - h^2)}, & \ep< -h.
    \end{cases} \label{CD}
\end{align} 

In the leading order in disorder strength, the current operator is modified by vertex corrections as
\begin{equation}
 \label{V}
 \mathbf{j}^\text{dress}
  = e\, (\mathbf{p} + \tilde{\alpha}\; \hat{\mathbf{z}} \times \bm{\sigma}),
 \qquad\qquad
 \tilde{\alpha}
  = \frac{1 + A/\tau}{1 - B/\tau}
  = \begin{cases}
      0, & \ep> h, \\
      1 - \dfrac{\lambda (\lambda_-^2 - h^2)}{\lambda_-^2 + 3 h^2}, & |\ep| < h, \\
      1 - \dfrac{\lambda^2}{1 + h^2}, & \ep< -h.
    \end{cases}
\end{equation}
The integrals $C$ and $D$ are not included in the vertex correction since they have a lower order in $\tau$. However, they are essential for the Hall response 
because they turn the $x$ component of the current into the $y$ component.

Below, we will also use the integral $D$ in the clean limit. Its computation amounts to setting $\eta$ to zero in Eq.\ (\ref{CD}). The intrinsic value 
of $D$ is
\begin{equation}
 D_0
  = \int \frac{ds}{2\pi}\; \left. \frac{2 \gamma h}{D^R(s) D^A(s)} \right|_{\eta = 0}
  = \frac{h}{2 \lambda} \begin{cases}
      \dfrac{\lambda}{\lambda^2 - 1}, & \ep> h, \\
      \dfrac{1}{2 \lambda_-}, & |\ep| < h, \\
      \dfrac{1}{1 - \lambda^2}, & \ep< -h.
    \end{cases} 
\end{equation}

The Hall conductivity $\sigma_{xy}^\text{I}$ can be expressed in the following form in the non-crossing approximation:
\begin{multline}
 \sigma_{xy}^\text{nc-I}
  = \frac{1}{2\pi} \int \frac{d^2p}{(2\pi)^2}\, \tr \bigl[ j_x^\text{dress} G^R(\mathbf{p}) j_y^\text{dress} G^A (\mathbf{p}) \bigr] \\
  = \frac{e^2}{2\pi} \int \frac{d^2p}{(2\pi)^2}\, \tr \bigl[
      (p_x - \tilde\alpha\, \sigma_y) G^R(\mathbf{p}) (p_y + \tilde\alpha\, \sigma_x) G^A (\mathbf{p})
    \bigr]
  = -\frac{e^2}{\pi} \tilde{\alpha} (2 C + \tilde{\alpha} D).
 \label{ncI}
\end{multline}
This result is given in Table \ref{table:nc} of the main text.

\subsection{Separation of intrinsic, side jump, and skew scattering terms}

In the non-crossing approximation, the Hall conductivity $\sigma^\text{nc-I}_{xy}$ is given by Eq.\ (\ref{ncI}). It can be represented as a sum of intrinsic 
($\sigma_{xy}^\text{int-I}$), side jump ($\sigma_{xy}^\text{side}$), and skew scattering ($\sigma_{xy}^\text{skew-nc}$) contributions as explained in 
the main text of the Letter. In order to develop such a classification, we first separate the conductivity $\sigma_{xy}^\text{nc}$ into three parts with zero, 
one, and two dressed current vertices. [By dressing we now assume only the leading order correction (\ref{V}), which does not rotate the direction of the 
current]. Relations between different contributions to $\sigma_{xy}$ are illustrated in Fig.\ \ref{Fig:sigma}. The three parts of $\sigma^\text{nc-I}_{xy}$ are
\begin{align}
 \sigma^\text{bare}_{xy}
  &= \frac{1}{2\pi} \int \frac{d^2p}{(2\pi)^2}\, \tr \bigl[ j_x G^R(\mathbf{p}) j_y G^A (\mathbf{p}) \bigr] \nonumber \\
  &= \frac{e^2}{2\pi} \int \frac{d^2p}{(2\pi)^2}\, \tr \bigl[ (p_x - \sigma_y) G^R(\mathbf{p}) (p_y + \sigma_x) G^A (\mathbf{p}) \bigr]
  = -\frac{e^2}{\pi} (2 C + D), \label{bare} \\
 \sigma^{1 \times \text{dress}}_{xy}
  &= \frac{1}{2\pi} \int \frac{d^2p}{(2\pi)^2}\, \tr \bigl[
      (j_x^\text{dress} - j_x) G^R(\mathbf{p}) j_y G^A (\mathbf{p})
      + j_x G^R(\mathbf{p}) (j_y^\text{dress} - j_y) G^A (\mathbf{p})
    \bigr] \nonumber \\
  &= \frac{e^2}{2\pi} (1 - \tilde\alpha) \int \frac{d^2p}{(2\pi)^2}\, \tr \bigl[
      2 \sigma_y G^R(\mathbf{p}) \sigma_x G^A (\mathbf{p})
      +\sigma_y G^R(\mathbf{p}) p_y G^A (\mathbf{p})
      -p_x G^R(\mathbf{p}) \sigma_x G^A (\mathbf{p})
    \bigr] \nonumber \\
  &= \frac{2e^2}{\pi} (1-\tilde{\alpha} )(C + D), \label{1dress} \\
 \sigma^{2 \times \text{dress}}_{xy}
  &= \frac{1}{2\pi} \int \frac{d^2p}{(2\pi)^2}\, \tr \bigl[
       (j_x^\text{dress} - j_x) G^R(\mathbf{p}) (j_y^\text{dress} - j_y) G^A (\mathbf{p})
     \bigr] \nonumber \\
  &= -\frac{e^2}{2\pi}(1 - \tilde\alpha)^2 \int \frac{d^2p}{(2\pi)^2}\, \tr \bigl[  \sigma_y G^R(\mathbf{p}) \sigma_x G^A (\mathbf{p}) \bigr]
  = -\frac{e^2}{\pi} (1-\tilde\alpha)^2 D.
\end{align}
Note that the part of the bare bubble diagram, $\sigma^\text{bare}_{xy}$, with two kinetic currents, $e p_x$ and $e p_y$, vanishes due to angular 
integration.

\begin{figure}
\centering
\renewcommand{\arraystretch}{1.4}
\newcolumntype{C}[1]{@{}>{\hsize=#1\hsize\centering\arraybackslash}X}
\begin{tabularx}{0.8\textwidth}{*8{|C{1}|}}
\hhline{--------}
\multicolumn{8}{|C{8}|}{$\sigma_{xy}$}\\
\hhline{=:t:=======}
$\sigma_{xy}^\text{II}$ & \multicolumn{7}{C{7}|}{$\sigma_{xy}^\text{I}$}\\
\hhline{=:b:=====:t:==}
\multicolumn{6}{|C{6}||}{$\sigma_{xy}^\text{nc}$} & \multicolumn{2}{C{2}|}{$\sigma_{xy}^{\text{X}+\Psi}$}\\
\hhline{=:t:=====::==}
$\sigma_{xy}^\text{II}$ & \multicolumn{5}{C{5}||}{$\sigma_{xy}^\text{nc-I}$} & \multicolumn{2}{C{2}|}{$\sigma_{xy}^{\text{X}+\Psi}$}\\
\hhline{=::==:t:==:t:=::=:t:=}
$\sigma_{xy}^\text{II}$ & \multicolumn{2}{C{2}||}{$\sigma_{xy}^\text{bare}$} & \multicolumn{2}{C{2}||}{$\sigma_{xy}^{1 \times \text{dress}}$}
& $\sigma_{xy}^{2 \times \text{dress}}$ & $\sigma_{xy}^\text{X}$ & $\sigma_{xy}^\Psi$ \\
\hhline{=::=:t:=:b:=:t:=:b:=::=::=}
$\sigma_{xy}^\text{II}$ & $\sigma_{xy}^\text{int-I}$ & \multicolumn{2}{C{2}||}{$\sigma_{xy}^\text{side}$} & 
\multicolumn{2}{C{2}||}{$\sigma_{xy}^\text{skew-nc}$} & $\sigma_{xy}^\text{X}$ & $\sigma_{xy}^\Psi$ \\
\hhline{=:b:=::==::==:b:=:b:=}
\multicolumn{2}{|C{2}||}{$\sigma_{xy}^\text{int}$} & \multicolumn{2}{C{2}||}{$\sigma_{xy}^\text{side}$}
& \multicolumn{4}{C{4}|}{$\sigma_{xy}^\text{skew}$}\\
\hhline{--||--||----}
\end{tabularx}
\caption{Different contributions to $\sigma_{xy}$ and relations between them.}
\label{Fig:sigma}
\end{figure}

Intrinsic, side jump, and skew scattering parts of $\sigma_{xy}$ are distinguished in the eigenbasis of the clean Hamiltonian. For each value of momentum 
$\mathbf{p}$ there are two eigenstates $|+\rangle$ and $|-\rangle$ corresponding to two branches of the spectrum. The Green function in the clean limit is 
diagonal in this basis, while disorder averaging gives rise to small off-diagonal terms $G_{+-}$ and $G_{-+}$. The current operator (\ref{j}) does not commute 
with the Hamiltonian and hence also possesses off-diagonal matrix elements in the eigenbasis. These off-diagonal terms emerge from the spin-orbit part of the 
current, $e\hat{\mathbf{z}} \times \bm{\sigma}$, while the kinetic part $e \mathbf{p}$ is diagonal. In the diagrammatic language, we identify intrinsic, side 
jump, and skew scattering contributions with the diagrams involving two, one, and none off-diagonal matrix elements of the currents, respectively.

Intrinsic Hall conductivity is the only part of total $\sigma_{xy}$ that survives in the clean limit. It is given by $\sigma_{xy}^\text{II}$ and the bare 
bubble (\ref{bare}) with clean Green functions. The latter is denoted $\sigma^\text{int-I}_{xy}$ and includes off-diagonal matrix elements of both current 
operators together with one $G_{++}$ and one $G_{--}$ element of the Green function.
\begin{equation}
 \sigma^\text{int-I}_{xy}
  = -\frac{e^2}{2\pi} \int \frac{d^2p}{(2\pi)^2}\, \tr \bigl[ \sigma_y G_0^R(\mathbf{p}) \sigma_x G_0^A (\mathbf{p}) \bigr]
  = -\frac{e^2}{\pi} D_0. 
\end{equation}

Side jump conductivity contains one diagonal and one off-diagonal matrix element of the current operator. Hence the rest of the bare bubble diagram (after 
subtracting the intrinsic contribution) yields side jump.
\begin{equation}
  \sigma^\text{bare}_{xy}
   = \underbrace{-\frac{e^2}{\pi} D_0}_\text{intrinsic} + \underbrace{\frac{e^2}{\pi} (D_0-2 C - D )}_\text{side jump}.
\end{equation}

Consider now the term $\sigma^{1 \times \text{dress}}_{xy}$. A part that contains the disorder-dressed current operator $\mathbf{j}^\text{dress} - \mathbf{j}$ 
connected to the bare current operator $\mathbf{j}$ with two clean Green functions, contributes to the side-jump conductivity. This intrinsic-like part of side 
jump is due to the spin-orbit part of the bare current only and comes completely from the first term in the second line of Eq.\ (\ref{1dress}). The rest of 
this term is distributed equally between side jump and skew scattering since both the dressed current and spin-orbit part of the bare current have the same 
off-diagonal matrix elements.
\begin{equation}
 \frac{e^2}{2\pi} (1 - \tilde\alpha) \int \frac{d^2p}{(2\pi)^2}\, \tr \bigl[ 2 \sigma_y G^R(\mathbf{p}) \sigma_x G^A (\mathbf{p}) \bigr]
  = \underbrace{\frac{e^2}{\pi} (1-\tilde{\alpha} ) (D + D_0)}_\text{side jump} + \underbrace{\frac{e^2}{\pi} (1-\tilde{\alpha} ) (D - D_0)}_\text{skew 
scattering}.
\end{equation}
The two other terms in the second line of Eq.\ (\ref{1dress}) yield skew scattering since the kinetic current $e \mathbf{p}$ does not provide off-diagonal 
matrix elements in the basis of the eigenstates. Thus we can summarize the contribution of $\sigma^{1 \times \text{dress}}_{xy}$ as
\begin{equation}
 \sigma^{1 \times \text{dress}}_{xy}
  = \underbrace{\frac{e^2}{\pi} (1-\tilde{\alpha} ) (D + D_0)}_\text{side jump}
    +\underbrace{\frac{e^2}{\pi} (1-\tilde{\alpha} ) (2C + D - D_0)}_\text{skew scattering}.
\end{equation}
Finally, $\sigma^{2 \times \text{dress}}_{xy}$ contributes solely to skew scattering mechanism,
\begin{equation}
 \sigma^{2 \times \text{dress}}_{xy}
  = \underbrace{-\frac{e^2}{\pi} (1-\tilde{\alpha} )^2 D}_\text{skew scattering}.
\end{equation}

Collecting together all the contributions, we obtain
\begin{align}
 \sigma^\text{int-I}_{xy}
  &= -\frac{e^2}{\pi} D_0, \\
 \sigma^\text{side}_{xy}
  &= \frac{e^2}{\pi} \bigl[ (2 - \tilde{\alpha}) D_0 - 2 C - \tilde{\alpha} D \bigr], \\
 \sigma^\text{skew-nc}_{xy}
  &= \frac{e^2}{\pi} (1 - \tilde{\alpha}) (2C + \tilde \alpha D - D_0).
\end{align}
These results are listed in the Table I of the main text.

\section{Calculation of X and $\bm{\Psi}$ diagrams}

An additional contribution to skew scattering comes from the diagrams with two intersecting impurity lines shown in Fig.\ \ref{fig:diag}b-d of the main text.
We will calculate these diagrams first in the momentum representation and later in the real-space representation as it is explained in the main text.

\subsection{Momentum representation}

The X diagram of Fig.\ \ref{fig:diag}b and $\Psi$ diagram of Fig.\ \ref{fig:diag}c-d are represented in momentum space by the following integrals: 
\begin{subequations}
\label{XPsi}
\begin{gather}
 \sigma_{xy}^\text{X}
  = \int \frac{d^2p_{1,2,3,4}}{(2\pi)^7 \tau^2}\, \delta(\mathbf{p}_1 + \mathbf{p}_2 - \mathbf{p}_3 - \mathbf{p}_4)
    \mathop{\mathrm{Tr}} \Bigl[
      j^\text{dress}_x G^R_1 G^R_3 G^R_2 j^\text{dress}_y G^A_2 G^A_4 G^A_1
    \Bigr], \label{X} \\
 \sigma_{xy}^{\rm{\Psi}}
  = \int \frac{d^2p_{1,2,3,4}}{(2\pi)^7 \tau^2}\, \delta(\mathbf{p}_1 - \mathbf{p}_2 - \mathbf{p}_3 + \mathbf{p}_4)
    \mathop{\mathrm{Tr}} \Bigl[
      j^\text{dress}_x G^R_1 G^R_3 G^R_4 G^R_2 j^\text{dress}_y G^A_2 G^A_1
      + j^\text{dress}_x G^R_1 G^R_2 j^\text{dress}_y G^A_2 G^A_4 G^A_3 G^A_1
    \Bigr]. \label{Psi}
\end{gather}
\end{subequations}

We first average the integrands with respect to simultaneous rotation of all momenta. This is equivalent to averaging with respect to rotations of the current 
operators, $j_x \mapsto j_x \cos\phi + j_y \sin\phi$ and $j_y \mapsto j_y \cos\phi - j_x \sin\phi$. We also use the symmetry with respect to $\mathbf{p}_{1,3} 
\leftrightarrow \mathbf{p}_{2,4}$ and rewrite the integrals as
\begin{gather}
 \sigma_{xy}^\text{X}
  = \int \frac{d^2p_{1,2,3,4}}{2 (2\pi)^7 \tau^2}\, \delta(\mathbf{p}_1 + \mathbf{p}_2 - \mathbf{p}_3 - \mathbf{p}_4)
    \frac{\mathop{\mathrm{Tr}} \bigl[
      J^x_1 N_3 J^y_2 N_4
    \bigr]}{D^R_1 D^A_1 D^R_2 D^A_2} \left( \frac{1}{D^R_3 D^A_4} - \frac{1}{D^A_3 D^R_4} \right), \\
 \sigma_{xy}^{\rm{\Psi}}
  = \int \frac{d^2p_{1,2,3,4}}{2 (2\pi)^7 \tau^2}\, \delta(\mathbf{p}_1 - \mathbf{p}_2 - \mathbf{p}_3 + \mathbf{p}_4)
    \frac{\mathop{\mathrm{Tr}} \bigl[
      (J^y_2 J^x_1 - J^x_2 J^y_1) N_3 N_4
    \bigr]}{D^R_1 D^A_1 D^R_2 D^A_2} \left( \frac{1}{D^R_3 D^R_4} - \frac{1}{D^A_3 D^A_4} \right),
\end{gather}
where we have introduced the short notation
\begin{equation}
 \mathbf{J}_i
  = N_i \mathbf{j}^\text{dress} N_i.
\end{equation}
We neglect disorder-induced $\gamma$ and $\eta$ terms in the numerators and hence make no distinction between $N^R$ and $N^A$. Next, we apply the identities
\begin{subequations}
\label{RRAA}
\begin{gather}
 \frac{1}{D^R_3 D^A_4} - \frac{1}{D^A_3 D^R_4}
  = \frac{1}{2} \left( \frac{1}{D^R_3} - \frac{1}{D^A_3} \right) \left( \frac{1}{D^R_4} + \frac{1}{D^A_4} \right)
    -\frac{1}{2} \left( \frac{1}{D^R_3} + \frac{1}{D^A_3} \right) \left( \frac{1}{D^R_4} - \frac{1}{D^A_4} \right), \\
 \frac{1}{D^R_3 D^R_4} - \frac{1}{D^A_3 D^A_4}
  = \frac{1}{2} \left( \frac{1}{D^R_3} - \frac{1}{D^A_3} \right) \left( \frac{1}{D^R_4} + \frac{1}{D^A_4} \right)
    +\frac{1}{2} \left( \frac{1}{D^R_3} + \frac{1}{D^A_3} \right) \left( \frac{1}{D^R_4} - \frac{1}{D^A_4} \right).
\end{gather}
\end{subequations}
Once again using the symmetry with respect to $\mathbf{p}_{1,3} \leftrightarrow \mathbf{p}_{2,4}$, we reduce the integrals to the form
\begin{gather}
 \sigma_{xy}^\text{X}
  = \int \frac{d^2p_{1,2,3,4}}{(2\pi)^7}\, \delta(\mathbf{p}_1 + \mathbf{p}_2 - \mathbf{p}_3 - \mathbf{p}_4)
    \frac{\mathop{\mathrm{Tr}} \bigl[
      J^x_1 N_3 J^y_2 N_4 - J^y_1 N_3 J^x_2 N_4
    \bigr]}{2 \tau^2 D^R_1 D^A_1 D^R_2 D^A_2} \left( \frac{1}{D^R_3} - \frac{1}{D^A_3} \right) \frac{1}{D_4}, \label{Xint}\\
 \sigma_{xy}^{\rm{\Psi}}
  = \int \frac{d^2p_{1,2,3,4}}{(2\pi)^7}\, \delta(\mathbf{p}_1 - \mathbf{p}_2 - \mathbf{p}_3 + \mathbf{p}_4)
    \frac{\mathop{\mathrm{Tr}} \bigl[
      (J^y_2 J^x_1 - J^x_2 J^y_1) N_3 N_4 + (J^y_1 J^x_2 - J^x_1 J^y_2) N_4 N_3
    \bigr]}{2 \tau^2 D^R_1 D^A_1 D^R_2 D^A_2} \left( \frac{1}{D^R_3} - \frac{1}{D^A_3} \right) \frac{1}{D_4}. \label{Psiint}
\end{gather}

From the identities (\ref{dendiff}) and (\ref{denden}) we see that momenta $\mathbf{p}_{1,2,3}$ are bound to the Fermi surface. To make use of this property,
we will employ the following double elliptic coordinates: $s_{1,2,3,4} = p^2_{1,2,3,4}/2$ and $s = (\mathbf{p}_1 + \mathbf{p}_2)^2/2 = (\mathbf{p}_3 + 
\mathbf{p}_4)^2/2$ for the X diagram or $s = (\mathbf{p}_1 - \mathbf{p}_2)^2/2 = (\mathbf{p}_3 - \mathbf{p}_4)^2/2$ for the ${\rm{\Psi}}$ diagram. The
integration measure in this representation is given by
\begin{gather} 
 \int \frac{d^2p_{1,2,3,4}}{(2\pi)^6}\; \delta(\mathbf{p}_1 \pm \mathbf{p}_2 - \mathbf{p}_3 \mp \mathbf{p}_4) \ldots
  = \int \frac{ds\, ds_{1,2,3,4}}{8 \pi^5 \sqrt{\Delta_{1,2}} \sqrt{\Delta_{3,4}}} \ldots, \\
 \Delta_{a,b}
  = -s^2 - s_a^2 - s_b^2 + 2 (s s_a + s s_b + s_a  s_b).
\end{gather}
Integral runs over the domain where both square roots in the denominator are real.

In order to represent the integrands in Eqs.\ (\ref{Xint}) and (\ref{Psiint}) as functions of $s$ variables, we further simplify the expressions by symmetrising 
them with respect to simultaneous flipping of $\mathbf{p}_{1,2}$ and/or $\mathbf{p}_{3,4}$ about the direction of $\mathbf{p}_1 \pm \mathbf{p}_2 = \mathbf{p}_3
\pm \mathbf{p}_4$; we will denote this operation by $\langle \ldots \rangle_P$. This results in
\begin{gather}
 \sigma_{xy}^{\mathrm{X}, {\rm{\Psi}}}
  = \int \frac{ds\, ds_{1,2,3,4}}{8 \pi^5 \sqrt{\Delta_{1,2}} \sqrt{\Delta_{3,4}}}\,
    \frac{N^{\mathrm{X}, {\rm{\Psi}}}_{1,2,3,4}}{\tau^2 D^R_1 D^A_1 D^R_2 D^A_2} \left( \frac{1}{D^R_3} - \frac{1}{D^A_3} \right) \frac{1}{D_4}, \\
 N^\mathrm{X}_{1,2,3,4}
  = \frac{1}{4 \pi}\mathop{\mathrm{Tr}} \Bigl<
      J^x_1 N_3 J^y_2 N_4 - \{ x \leftrightarrow y \}
    \Bigr>_P,
 \qquad
 N^{\rm{\Psi}}_{1,2,3,4}
  = \frac{1}{4 \pi}\mathop{\mathrm{Tr}} \Bigl<
      J^y_2 J^x_1 N_3 N_4 + J^y_1 J^x_2 N_4 N_3 - \{ x \leftrightarrow y \}
    \Bigr>_P.
\end{gather}
We will need the value of the averaged numerators $N^{\mathrm{X}, {\rm{\Psi}}}_{1,2,3,4}$ at the points $s_{1,2} = s_\pm$ only. An explicit computation of the
trace yields remarkably simple expressions:
\begin{align}
 N^\mathrm{X}_{\pm,\pm,3,4}
  &= -\frac{2 i e^2 h}{\pi s} \Delta_{\pm\pm} (\tilde{\alpha} \pm \lambda_\pm)^2 (s_3 - s_4) (s_3 + s_4 - 2 s_\mp + 4), \\
 N^\mathrm{X}_{\pm,\mp,3,4}
  &= -\frac{2 i e^2 h}{\pi s} \Delta_{+-} (\tilde{\alpha} + \lambda_+) (\tilde{\alpha} - \lambda_-) (s_3 - s_4) (s_3 + s_4 - s_+ - s_- + 4), \\
 N^{\rm{\Psi}}_{\pm,\pm,3,4}
  &= \frac{4 i e^2 h}{\pi} \Delta_{\pm\pm} (\tilde{\alpha} \pm \lambda_\pm)^2 (s_3 + s_4 - 2 s_\mp + 4), \\
 N^{\rm{\Psi}}_{\pm,\mp,3,4}
  &= \frac{4 i e^2 h}{\pi} \Delta_{+-} (\tilde{\alpha} + \lambda_+) (\tilde{\alpha} - \lambda_-) (s_3 + s_4 - s_+ - s_- + 4).
\end{align}

At a given value of the variable $s$, and with $s_{1,2,3}$ fixed by the delta functions from Eqs.\ (\ref{dendiff}) and (\ref{denden}), the integral over $s_4$ 
is taken in the interval $(\sqrt{s} - \sqrt{s_3})^2 < s_4 < (\sqrt{s} + \sqrt{s_3})^2$. We can represent this integral as a contour integral around the branch 
cut of $\sqrt{-\Delta_{3,4}}$ and then expand the contour to an infinite circle. This way we pick the residues at the poles of $1/D_4$, if they lie outside the 
branch cut, and the residue at infinity. In the case $|\ep| < h$, this also includes the residue at $s_4 = s_+ < 0$. The contour transform can be presented as
\begin{multline}
 \int\limits_{(\sqrt{s} - \sqrt{s_3})^2}^{(\sqrt{s} + \sqrt{s_3})^2} \frac{ds_4}{\sqrt{\Delta_{3,4}}\, D_4} \ldots
  = -\pi \sum \mathop{\mathrm{res}}_{s_4} \frac{1}{\sqrt{-\Delta_{3,4}}\, (s_4 - s_-)(s_4 - s_+)} \ldots \\
  = \frac{\pi}{2\lambda} \int \frac{ds_4}{\sqrt{-\Delta_{3,4}}}
      \Bigl[ \delta(s_4 - s_+) - \delta(s_4 - s_-) \Bigr] \mathop{\mathrm{sign}} (s_4 - s_3 - s) \ldots
    +\pi \lim_{s_4 \to \infty} \frac{1}{s_4^2} \ldots
\end{multline}
Here the sign factor accounts for a proper branch of the square root while the expression $\sqrt{-\Delta_{3,4}}$ assumes the principal value. This way we 
effectively fix all four momenta at the Fermi surface with the additional possibility $s_4 = \infty$ and a constraint $s_{1,2,3} > 0$. The only remaining 
integration over $s$, as before, runs over positive real semi-axis such that both $\sqrt{\Delta_{1,2}}$ and $\sqrt{-\Delta_{3,4}}$ are real.

After quite tedious combinatorics, we identify all integration intervals for $s$, which contribute to X and ${\rm{\Psi}}$ diagrams, depending on the value of
energy. They are listed in the Table \ref{table:regions}.
\begingroup
\begin{table}
\begin{center}
\tabcolsep=0.5em
\begin{tabular}{cccccccccccc}
\hline\hline\\[-8pt]
  \raisebox{-6pt}[6pt][0pt]{Label} &
  \raisebox{-6pt}[6pt][0pt]{$s_1$} &
  \raisebox{-6pt}[6pt][0pt]{$s_2$} &
  \raisebox{-6pt}[6pt][0pt]{$s_3$} &
  \raisebox{-6pt}[6pt][0pt]{$s_4$} &
  \raisebox{-6pt}[6pt][0pt]{Domain for $s$} &
  \multicolumn{2}{c}{$\ep> h$} &
  \multicolumn{2}{c}{$|\ep| < h$} & 
  \multicolumn{2}{c}{$\ep< -h$} \\[2pt]
\cline{7-12}\\[-8pt]
&&&&&&
  \hspace{0.3em} X \hspace{0.3em} & \hspace{0.3em} ${\rm{\Psi}}$ \hspace{0.3em} &
  \hspace{0.3em} X \hspace{0.3em} & \hspace{0.3em} ${\rm{\Psi}}$ \hspace{0.3em} &  
  \hspace{0.3em} X \hspace{0.3em} & \hspace{0.3em} ${\rm{\Psi}}$ \hspace{0.3em} \\
\hline\\[-6pt]
A & $+$ & $+$ & $\pm$ & $\mp$ & $\bigl[ 0, \min\{ (\sqrt{s_-} - \sqrt{s_+})^2, 4 s_+\} \bigr]$ & \checkmark &&&&& \checkmark \\[3pt]
B & $\pm$ & $\mp$ & $+$ & $+$ & $\bigl[ \max\{(\sqrt{s_-} - \sqrt{s_+})^2, 4 s_+ \}, (\sqrt{s_-} + \sqrt{s_+})^2 \bigr]$ && \checkmark &&&& \checkmark\\[3pt]
C & $-$ & $-$ & $+$ & $+$ & $[4 s_+, 4 s_-]$ && \checkmark &&&& \checkmark \\[3pt]
D & $-$ & $-$ & $\pm$ & $\mp$ & $[0, (\sqrt{s_-} - \sqrt{s_+})^2]$ & \checkmark &&&&& \checkmark \\[3pt]
E & $-$ & $-$ & $\pm$ & $\mp$ & $[(\sqrt{s_-} + \sqrt{s_+})^2, 4 s_-]$ && \checkmark &&& \checkmark \\[3pt]
DE & $-$ & $-$ & $-$ & $+$ & $[0, 4 s_-]$ &&& \checkmark & \checkmark \\[3pt]
F & $+$ & $+$ & $\forall$ & $\infty$ & $[0, 4s_+]$ &&&&& \checkmark \\[3pt]
G & $\pm$ & $\mp$ & $\forall$ & $\infty$ & $[(\sqrt{s_-} - \sqrt{s_+})^2, (\sqrt{s_-} + \sqrt{s_+})^2]$ &&&&& \checkmark \\[3pt]
H & $-$ & $-$ & $\forall$ & $\infty$ & $[0, 4s_-]$ &&& \checkmark && \checkmark \\[3pt]
\hline\hline
\end{tabular}
\end{center}
\caption{\label{table:regions}
Integration intervals for the parameter $s$ that contribute to X and $\rm{\Psi}$ diagrams for different parts of the spectrum.}
\end{table}
\endgroup

Note that inside the gap, $|\ep| < h$, there is only one Fermi surface at $s = s_-$ while $s_+ < 0$. In this case the intervals D and E fuse into a single 
segment DE. The intervals F, G, and H contribute when the residue at $s_4 = \infty$ is taken and the value of $s_3$ is unimportant.

Explicitly, the diagrams with crossed impurity lines yield
\begin{align}
 \sigma_{xy}^\mathrm{X}
  = -\frac{e^2 h \tau_\text{tr}^2}{4 \pi^2 \lambda^3 \tau^2}
    &\begin{dcases}
      (2 - \lambda) \int\limits_A \frac{ds}{s}\, \frac{\sqrt{\Delta_{++}}}{\sqrt{-\Delta_{+-}}}
      +(2 + \lambda) \int\limits_D \frac{ds}{s}\, \frac{\sqrt{\Delta_{--}}}{\sqrt{-\Delta_{+-}}},
        & \ep> h, \\
      \left( 1 + \frac{\lambda}{2} \right) \int\limits_{DE} \frac{ds}{s}\, \frac{\sqrt{\Delta_{--}}}{\sqrt{-\Delta_{+-}}}
      +\int\limits_H \frac{ds}{4 s}\, \sqrt{\Delta_{--}}
        & |\ep| < h, \\
      (2 + \lambda) \int\limits_E \frac{ds}{s}\, \frac{\sqrt{\Delta_{--}}}{\sqrt{-\Delta_{+-}}}
      +\int\limits_F \frac{ds}{2s}\, \sqrt{\Delta_{++}}
      -\int\limits_G \frac{ds}{s}\, \sqrt{\Delta_{+-}}
      +\int\limits_H \frac{ds}{2s}\, \sqrt{\Delta_{--}},
        & \ep< -h, 
    \end{dcases} \label{Xresult} \\
 \sigma_{xy}^{\rm{\Psi}}
  = \frac{e^2 h \tau_\text{tr}^2}{4 \pi^2 \lambda^4 \tau^2}  
    &\begin{dcases}
      (2 - \lambda)\int\limits_B \frac{ds \sqrt{\Delta_{+-}}}{\sqrt{-\Delta_{++}}}
      -\int\limits_C \frac{ds \sqrt{\Delta_{--}}}{\sqrt{-\Delta_{++}}}
      +(2 + \lambda) \int\limits_E\frac{ds \sqrt{\Delta_{--}}}{\sqrt{-\Delta_{+-}}},
        & \ep> h, \\
      \left( 1 + \frac{\lambda}{2} \right) \int\limits_{DE} \frac{ds \sqrt{\Delta_{--}}}{\sqrt{-\Delta_{+-}}}, & |\ep| < h, \\
      (2 - \lambda) \left(
        \int\limits_A \frac{ds \sqrt{\Delta_{++}}}{\sqrt{-\Delta_{+-}}}
        -\int\limits_B \frac{ds \sqrt{\Delta_{+-}}}{\sqrt{-\Delta_{++}}}
      \right)
      +\int\limits_C \frac{ds \sqrt{\Delta_{--}}}{\sqrt{-\Delta_{++}}}
      +(2 + \lambda) \int\limits_D \frac{ds \sqrt{\Delta_{--}}}{\sqrt{-\Delta_{+-}}}
        & \ep< -h.
    \end{dcases} \label{Psiresult}
\end{align}
We use the transport scattering time $\tau_\text{tr}$ defined in the next section, see Eq.\ (\ref{ttr}). Let us also remind that
\begin{equation}
 \Delta_{--}
  = s (4 s_- - s),
 \qquad
 \Delta_{++}
  = s (4 s_+ - s),
 \qquad
 \Delta_{+-}
  = -s^2 + 2 s (s_- + s_+) - (s_- - s_+)^2.
\end{equation}
The integrals over the intervals C, F, G, and H are readily computed in terms of elementary functions while the intervals A, B, D, E (and DE) contain complete 
elliptic integrals. Computation of these integrals is discussed in Appendix A. Final results are given after the following Section and also in the main text.

\subsection{Real space representation}

Alternatively, X and ${\rm{\Psi}}$ diagrams can be calculated in the real space representation. This way of calculation is more transparent but leads 
to a large variety of integrals involving four Bessel functions. 

The Green's function in real space can be decomposed in the clean limit, $\tau\to \infty$, into two contributions originating from two branches of the spectrum,
\begin{gather}
\label{realGreens1}
 G^{R,A}(\mathbf{r})
  = G^{R,A}_+(\mathbf{r}) + G^{R,A}_-(\mathbf{r}), \\
  \label{realGreens2}
 G^{R,A}_\pm(\mathbf{r})
  = \pm \frac{1}{2 \lambda}\, \bigl( \ep+ \nabla^2/2 + i (\bm{\sigma} \times \nabla)_z + h \sigma_z \bigr)\, g^{R,A}_\pm(r), \\
  \label{realGreens3}
 \begin{aligned}
 g^{R,A}_-(r)
  &= \int \frac{d^2 p}{(2\pi)^2} \frac{e^{i \mathbf{p} \mathbf{r}}}{s_- - p^2/2 \pm i \gamma_-}
  = \frac{1}{2} \bigl[ Y_0(p_- r) \mp i J_0 (p_- r) \bigr], \\
 g^{R,A}_+(r)
  &= \int \frac{d^2 p}{(2\pi)^2} \frac{e^{i \mathbf{p} \mathbf{r}}}{s_+ - p^2/2 \pm i \gamma_+}
  = \begin{dcases}
      \frac{1}{2} \bigl[ Y_0(p_+ r) \mp i \mathop{\mathrm{sign}} \gamma_+ J_0 (p_+ r) \bigr], & |\ep| > h, \\
      -\frac{1}{\pi} K_0(|p_+| r), & |\ep| < h,
    \end{dcases}
 \end{aligned}
\end{gather}
where we use the notation $p_\pm = \sqrt{2 s_\pm}$ for the two Fermi momenta. Here we completely neglect the values of $\gamma_\pm$ by taking them as infinitesimals. Note that at $|\ep| < h$, when only $p_-$ Fermi surface exists, there is no distinction between $G^R_+$ and $G^A_+$. Both functions decay monotonically with  distance.

Apart from the Green's function, we also need the real space representation of the dressed current vertex (in the leading order with respect to $1/\tau$.)
\begin{multline}
 \mathbf{J}(\mathbf{r})
  = e \int \frac{d^2 p}{(2\pi)^2}\, e^{i \mathbf{p} \mathbf{r}}\,
      G^R(\mathbf{p}) (\mathbf{p} + \tilde{\alpha} \hat{\mathbf{z}} \times \bm{\sigma}) G^A(\mathbf{p}) \\
  = \frac{ie}{8\lambda^2}
    \bigl[ \ep + \nabla^2/2 + i (\bm{\sigma} \times \nabla)_z + h \sigma_z \bigr]
    \bigl( -i \nabla + \tilde{\alpha} \hat{\mathbf{z}} \times \bm{\sigma} \bigr)
    \bigl[ \ep + \nabla^2/2 + i (\bm{\sigma} \times \nabla)_z + h \sigma_z \bigr] \\
    \times \left[ \frac{g_-^R(r) - g_-^A(r)}{\gamma_-} + \frac{g_+^R(r) - g_+^A(r)}{\gamma_+} \right],
\end{multline}
where the factor $\tilde{\alpha}$ accounts for the vertex correction (\ref{V}). Below we take advantage of the following two identities
\begin{align}
 &-i \nabla \bigl[ \ep + \nabla^2/2 + i (\bm{\sigma} \times \nabla)_z + h \sigma_z \bigr]^2 \notag \\
    &\hspace{4cm} = -i \nabla \biggl\{
       \underbrace{-(\ep + \nabla^2/2)^2 - \nabla^2 + h^2}_{= 0} + 2(\ep+ \nabla^2/2) \bigl[ \ep+ \nabla^2/2 + i (\bm{\sigma} \times \nabla)_z + h \sigma_z \bigr]
     \biggr\}, \\
 &\bigl[ \ep+ \nabla^2/2 + i (\bm{\sigma} \times \nabla)_z + h \sigma_z \bigr]
    (\hat{\mathbf{z}} \times \bm{\sigma})
  \bigl[ \ep+ \nabla^2/2 + i (\bm{\sigma} \times \nabla)_z + h \sigma_z \bigr] \notag \\
   &\hspace{4cm} = \underbrace{\bigl[ (\ep+ \nabla^2/2)^2 + \nabla^2 - h^2 \bigr]}_{= 0} (\hat{\mathbf{z}} \times \bm{\sigma})
    -2 i \nabla \bigl[ \ep+ \nabla^2/2 + i (\bm{\sigma} \times \nabla)_z + h \sigma_z \bigr],
\end{align}
where the action of the operator $(\ep+ \nabla^2/2)^2 + \nabla^2 - h^2$ on $g_\pm^{R,A}$ yields zero.  With the help of the identities we are able to recast 
the current operator in the following form
\begin{multline}
 \mathbf{J}(\mathbf{r})
  = \frac{e \nabla}{4 \lambda^2}
    (\ep+ \nabla^2/2 + \tilde{\alpha}) \bigl( \ep+ \nabla^2/2 - i \bm{\sigma}\nabla + h \sigma_z \bigr)
    \left[ \frac{g_-^R(r) - g_-^A(r)}{\gamma_-} + \frac{g_+^R(r) - g_+^A(r)}{\gamma_+} \right] \\
  = \frac{e \nabla}{2 \lambda} \left[
      \frac{\lambda_- - \tilde{\alpha}}{\gamma_-} \bigl( G_-^R(\mathbf{r}) - G_-^A(\mathbf{r}) \bigr)
      +\frac{\lambda_+ + \tilde{\alpha}}{\gamma_+} \bigl( G_+^R(\mathbf{r}) - G_+^A(\mathbf{r}) \bigr)
    \right].
    \label{multipulti}
\end{multline}
Using Eq.\ (\ref{V}), we see that for an energy outside the gap, $|\ep| > h$, two factors in the last expression are identical up to a sign. Inside the gap,
$|\ep| < h$, the second term in Eq.~(\ref{multipulti}) is unimportant since the difference $G_+^R - G_+^A$ vanishes. This allows us to introduce the transport
scattering time $\tau_\text{tr}$ and rewrite the current vertex in a remarkably compact form:
\begin{equation}
 \mathbf{J}(\mathbf{r})
  = e \tau_\text{tr} \nabla \Bigl[
      G_-^R(\mathbf{r}) - G_-^A(\mathbf{r}) + \mathop{\mathrm{sign}} \ep \Bigl( G_+^R(\mathbf{r}) - G_+^A(\mathbf{r}) \Bigr)
    \Bigr],
 \qquad
 \tau_\text{tr}
  = \frac{\lambda_\pm \pm \tilde{\alpha}}{2\lambda |\gamma_\pm|}
  = \tau \begin{dcases}
      1, & \ep> h, \\
      \frac{4 \lambda^2}{\lambda_-^2 + 3 h^2}, & |\ep| < h, \\
      \frac{\lambda^2}{1 + h^2}, & \ep< -h.
    \end{dcases} \label{ttr}
\end{equation}
From this expression we see that electrons from both spectrum branches provide identical contributions to the current. The only caveat is that the sign of $+$
branch changes when energy drops below the gap, $\ep< -h$. This is a manifestation of the hole-like dispersion of the $+$ part of the spectrum in this region.
It is also worth noting, that the transport time $\tau_\text{tr}$ is introduced here phenomenologically and does not necessarily correspond to any specific rate
derived from kinetic 
equation. The latter has a complicated form whenever two branches of the spectrum are present, hence physical scattering rates will have a certain matrix
structure and may not boil down to a single parameter.

Expressions for the X and $\Psi$ diagrams in the real space representation are given by Eqs.\ (\ref{XY}) of the main text. Using cyclic permutations under the 
trace, the symmetry relation $\sigma_{xy} = -\sigma_{yx}$, and identities of the type of Eqs.\ (\ref{RRAA}), we can recast them in the form
\begin{gather}
 \sigma_{xy}^{\rm{X}}
  = \frac{1}{4\pi} \frac{1}{\tau^2} \int d^2r \mathop{\mathrm{Tr}} \Bigl\{
      \bigl( G^A_\mathbf{r} + G^R_\mathbf{r} \bigr) J^x_{-\mathbf{r}} \bigl( G^R_\mathbf{r} - G^A_\mathbf{r} \bigr) J^y_{-\mathbf{r}}
    \Bigr\}, \label{Xreal}\\
 \sigma_{xy}^{\rm{\Psi}}
  = -\frac{1}{4\pi} \frac{1}{\tau^2} \int d^2r \mathop{\mathrm{Tr}} \Bigl\{
      \bigl( G^R_\mathbf{r} + G^A_\mathbf{r} \bigr) \Bigl[
        J^x_{-\mathbf{r}} J^y_\mathbf{r} \bigl( G^R_{-\mathbf{r}} - G^A_{-\mathbf{r}} \bigr)
        +\bigl( G^R_{-\mathbf{r}} - G^A_{-\mathbf{r}} \bigr) J^x_{-\mathbf{r}} J^y_\mathbf{r}
      \Bigr]
    \Bigr\}. \label{Psireal}
\end{gather}

After averaging over directions of $\mathbf{r}$, we end up with a large number of one-dimensional integrals involving four Bessel functions. Current operators 
and the difference $G^R - G^A$ yield the Bessel function of the first kind $J_\nu(p_\pm r)$, while the sum $G^R + G^A$ provides either $Y_\nu(p_\pm r)$ or 
$K_\nu(|p_+| r)$ in the regions $|\ep| > h$ and $|\ep| < h$, respectively. For the sake of convenience, we will use the following compact notations:
\begin{equation}
 J_\nu^{\pm} \equiv p_\pm^\nu J_\nu (p_{\pm} r), \qquad Y_\nu^{\pm} \equiv p_\pm^\nu Y_\nu (p_{\pm} r), \qquad K_\nu^{+} \equiv |p_+|^\nu K_\nu (|p_{+}| r).
 \label{compact}
\end{equation}

With the help of recurrence relations, we reduce orders of all Bessel functions to either $\nu = 0$ or $\nu = 1$ for uniformity. For $|\varepsilon| > h$ the 
result is given by a sum of $34$ (in the case of $\sigma_{xy}^{\mathrm{X}}$) and $44$ (in the case of $\sigma_{xy}^{\rm{\Psi}}$) distinct integrals of four 
Bessel functions. For $|\varepsilon| < h$, the expressions are more compact and can be represented by the sum of $6$ and $8$ integrals, correspondingly. 

Apart from four Bessel functions, some terms in the integrand contain an extra factor $1/r$. It turns out that the formal replacement $1/r \mapsto 1/r + (1/2) \partial/\partial r$ eliminates all $1/r$ terms and renders the integrand uniform. 
Let us illustrate the mechanism of such a reduction by the following example:
\begin{equation}
 \int\limits_0^\infty \frac{dr}{r}\, Y_0^- (J_1^-)^2 J_0^+
  = \int\limits_0^\infty dr \left( \frac{1}{r} + \frac{1}{2}\, \frac{\partial}{\partial r} \right)\, Y_0^- (J_1^-)^2 J_0^+
  = \int\limits_0^\infty dr \left[
      p_-^2 Y_0^- J_0^- J_1^- J_0^+ - \frac{1}{2}\, Y_1^- (J_1^-)^2 J_0^+ - \frac{1}{2}\, Y_0^- (J_1^-)^2 J_1^+
    \right].
\end{equation}
After reduction, we end up with $20$ distinct integrals for $|\varepsilon| > h$ and two additional integrals for $|\varepsilon| < h$. The expressions for 
$\sigma_{xy}^\text{X}$ and $\sigma_{xy}^\Psi$ can be concisely written using the following matrix notations:
\begin{gather}
 \sigma_{xy}^\text{X}
  = \frac{e^2 h \tau_\text{tr}^2}{8 \lambda^4 \tau^2} \int\limits_0^\infty dr\, [X_{10} - X_{01}]\, R\, Y^T,
 \qquad\qquad
 \sigma_{xy}^\Psi
  = \frac{e^2 h \tau_\text{tr}^2}{4 \lambda^4 \tau^2} \int\limits_0^\infty dr\, [X_{10} + X_{01}]\, R\, Y^T, \label{XYBessel} \\
 R
  = \begin{pmatrix}
      1 & 1 + \lambda/2 & 1 + \lambda \\
      1 - \lambda/2 & 1 & 1 + \lambda/2 \\
      1 - \lambda & 1 - \lambda/2 & 1
    \end{pmatrix}, \\
 X_{\mu\nu}
  = \begin{cases}
      \begin{pmatrix} Y_\mu^- J_\nu^-, & -Y_\mu^- J_\nu^+ - Y_\mu^+ J_\nu^-, & Y_\mu^+ J_\nu^+ \end{pmatrix}, & \ep > h, \\
      \begin{pmatrix} Y_\mu^- J_\nu^-, & (2/\pi) K_\mu J_\nu^-, & 0 \end{pmatrix}, & |\ep| < h, \\
      \begin{pmatrix} Y_\mu^- J_\nu^-, & Y_\mu^- J_\nu^+ - Y_\mu^+ J_\nu^-, & -Y_\mu^+ J_\nu^+ \end{pmatrix}, & \ep < -h,
    \end{cases}
 \qquad
 Y
  = \begin{cases}
      \begin{pmatrix} (J_1^+)^2, & -2 J_1^+ J_1^-, & (J_1^-)^2 \end{pmatrix}, & |\ep| > h, \\
      \begin{pmatrix} 0, & 0, & (J_1^-)^2 \end{pmatrix}, & |\ep| < h.
    \end{cases}
\end{gather}

Thus we have reduced the problem to a set of integrals involving three Bessel functions of the first kind and one Bessel function of the second kind with 
arguments $p_\pm r$. One out of four Bessel functions has the index $0$ and three other functions bear the index $1$. Calculation of these integrals is detailed 
in Appendix B, and the result is given in the next Section.

\subsection{Results}

The two approaches to the X and $\rm{\Psi}$ diagram in momentum and in real space yield the same result, which makes us confident that the calculation is 
accurate. The final expressions for the diagrams can be conveniently written in terms of the auxiliary parameter
\begin{equation}
 a
  = \begin{dcases}
      \frac{\sqrt{s_-} - \sqrt{s_+}}{\sqrt{s_-} + \sqrt{s_+}}, & |\ep| > h, \\
      \frac{1}{2} \left[ \sqrt{\frac{9 s_- - s_+}{s_- - s_+}} - 1 \right], & |\ep| < h.
    \end{dcases} \label{a}
\end{equation}
This parameter is in the range $0 < a < 1$ taking the value $1$ at $\ep= \pm h$ and vanishing in the limit of infinite energy and at the bottom of the band
$\ep = -(1 + h^2)/2$. Complete elliptic integrals arising in Eqs.\ (\ref{Xresult}) and (\ref{Psiresult}), have one of the following moduli:
\begin{equation}
 k_\pm^2
  = \frac{a^3 (2 \pm a)}{2a \pm 1},
 \qquad
 {k'_\pm}^2
  = 1 - k_\pm^2
  = \frac{(1 \pm a)^3 (1 \mp a)}{1 \pm 2a}.
 \label{moduli}
\end{equation}
We will use the short notations $K_\pm = K(k_\pm)$, $K'_\pm = K(k'_\pm)$ for the complete elliptic integrals of the first kind and the similar abbreviation 
$E_\pm$, $E'_\pm$ for the complete integrals of the second kind.

For all energies $0 < k_+ < 1$. The other module is either $k_-^2 < 0$ when $0 < a < 1/2$ or $k_-^2 >1$ when $1/2 < a < 1$. Switching between these two cases 
occurs when the energy takes one of the values
\begin{equation}
 \epsilon_\pm  = \frac{9 \pm 5 \sqrt{9 + 16 h^2}}{16}.
\end{equation}
These special values of energy correspond to the perfect ``nesting'' configuration $\sqrt{s_-} = 3 \sqrt{s_+}$ when the minimum and maximum in the definition 
of $A$ and $B$ intervals change. Such nesting occurs once (at $\ep= \epsilon_+$) above the gap and once (at $\ep= \epsilon_-$) below the gap.

Thus we have in total five intervals of energy where the expression for Hall conductivity acquires different functional form. Using the results of Appendix A 
or B, we find 
\begin{equation}
 \sigma_{xy}^\mathrm{X + \rm{\Psi}}
  = \frac{e^2 h}{2 \pi^2} \begin{dcases}
      \frac{1}{\lambda^3 a} \Bigl[
        (2 + \lambda) \sqrt{1 + 2a} (K_+ - E_+) + (2 - \lambda) \sqrt{1 - 2a} (K_- - E_-)
      \Bigr], & \ep> \epsilon_+, \\
      \frac{1}{\lambda^3 a} \Bigl[
        (2 + \lambda) \sqrt{1 + 2a} (K_+ - E_+) - (2 - \lambda) \sqrt{2a - 1} E'_-
      \Bigr], & h < \ep< \epsilon_+, \\
      \frac{4 \lambda}{(\lambda_-^2 + 3 h^2)^2} \Bigl[
        (2 + \lambda) \bigl[ \pi + 2 \sqrt{1 + 2a} (K_+ - 2 E_+) \bigr] - \pi \lambda a (1 + a)
      \Bigr], & -h < \ep< h, \\
      \frac{\lambda}{(1 + h^2)^2} \Bigl[
        (2 + \lambda) \frac{\sqrt{1 + 2a}}{a} (K_+ - E_+) + \frac{2}{3} (2 - \lambda) \bigl[ \pi + (2 - a) \sqrt{2a - 1} K'_- \bigr]
      \Bigr], & \epsilon_- < \ep< -h, \\
      \frac{\lambda}{(1 + h^2)^2} \Bigl[
        (2 + \lambda) \frac{\sqrt{1 + 2a}}{a} (K_+ - EK_+) + \frac{2}{3} (2 - \lambda) \bigl[ \pi - (2 - a) \sqrt{1 - 2a} K_- \bigr]
      \Bigr], & \epsilon_\text{min}<\ep< \epsilon_-.
    \end{dcases}
 \label{result}
\end{equation}
This is the central result of the Letter presented in the Table \ref{table:result} of the main text. When the parameter $h$ exceeds $1$ (that is, 
ferromagnetism is stronger than spin-orbit coupling) the structure of the spectrum changes. Minimal available  energy becomes $-h$ and only the first three 
out of five cases in Eq. (\ref{result}) remain.

\appendix

\subsection*{Appendix A: Elliptic integrals}

All elliptic integrals from Eqs.\ (\ref{Xresult}) and (\ref{Psiresult}) can be found in Ref.\ \onlinecite{Byrd} in the incomplete form. They involve elliptic 
integrals of all three kinds with the moduli (\ref{moduli}). Elliptic integrals of the third kind arising in this calculation depend on the parameter $-a^2/(2a
\pm 1)$. However, the complete versions of the same integrals can be written in terms of the first and second kind integrals only. This is possible due to the
following identity:
\begin{equation}
\label{magic}
 \Pi \left(
   -\frac{a^2}{2a + 1}, k_+
 \right)
  = \frac{\pi \sqrt{2a + 1} + 2(2 + 5 a + 2 a^2) K(k_+)}{6 (1 + a)^2}.
\end{equation}

Outside the gap, $|\ep| > h$, the relevant integrals have the following form:
\begin{align}
 &\int_A \frac{ds}{s}\, \frac{\sqrt{\Delta_{++}}}{\sqrt{-\Delta_{+-}}}
  = \frac{\pi}{3} + \frac{2}{3} (2 - 5 a + 2 a^2) \begin{dcases}
      \frac{K_-}{\sqrt{1 - 2a}}, & 0 < a < 1/2, \\
      \frac{K'_-}{\sqrt{2a - 1}}, & 1/2 < a < 1,
    \end{dcases} \\
 &\int_A \frac{ds}{\lambda}\, \frac{\sqrt{\Delta_{++}}}{\sqrt{-\Delta_{+-}}}
  = \frac{2 \pi}{3} - \frac{2}{a} \begin{dcases}
      \sqrt{1 - 2a} \left[ E_- - \frac{3 - 2a + a^2}{3} K_- \right], & 0 < a< 1/2, \\
      \sqrt{2a - 1} \left[ E'_- - \frac{(2 - a) a}{3} K'_- \right], & 1/2 < a < 1,
    \end{dcases} \\
 &\int_B \frac{ds}{\lambda}\, \frac{\sqrt{\Delta_{+-}}}{\sqrt{-\Delta_{++}}}
  = \frac{4 \pi}{3} - \frac{2}{a} \begin{dcases}
      \sqrt{1 - 2a} \left[ E_- - \frac{(3 - a)(1 + a)}{3} K_- \right], & 0 < a < 1/2, \\
      \sqrt{2a - 1} \left[ E'_- + \frac{(2 - a) a}{3} K'_- \right], & 1/2 < a < 1
    \end{dcases} \\
 &\int_C \frac{ds}{\lambda}\, \frac{\sqrt{\Delta_{--}}}{\sqrt{-\Delta_{++}}}
  = 4 \pi, \\
 &\int_D \frac{ds}{s}\, \frac{\sqrt{\Delta_{--}}}{\sqrt{-\Delta_{+-}}}
  = \frac{\pi}{3} + \frac{2(2 + 5 a + 2 a^2)}{3 \sqrt{2a + 1}} K_+, \\
 &\int_D \frac{ds}{\lambda}\, \frac{\sqrt{\Delta_{--}}}{\sqrt{-\Delta_{+-}}}
  = -\frac{2 \pi}{3} - \frac{2 \sqrt{2a + 1}}{a} \left[ E_+ - \frac{3 + 2a + a^2}{3} K_+ \right], \\
 &\int_E \frac{ds}{s}\, \frac{\sqrt{\Delta_{--}}}{\sqrt{-\Delta_{+-}}}
  = -\frac{2\pi}{3} + \frac{2(2 + 5 a + 2 a^2)}{3 \sqrt{2a + 1}} K_+, \\
 &\int_E \frac{ds}{\lambda}\, \frac{\sqrt{\Delta_{--}}}{\sqrt{-\Delta_{+-}}}
  = \frac{4 \pi}{3} - \frac{2 \sqrt{2a + 1}}{a} \left[ E_+ - \frac{3 + 2a + a^2}{3} K_+ \right], \\
 &\int_F \frac{ds}{\lambda s}\, \sqrt{\Delta_{++}}
  = \frac{\pi (1 - a)^2}{a}, \\
 &\int_G \frac{ds}{\lambda s}\, \sqrt{\Delta_{+-}}
  = \frac{\pi (1 - a)^2}{a}, \\
 &\int_H \frac{ds}{\lambda s}\, \sqrt{\Delta_{--}}
  = \frac{\pi (1 + a)^2}{a}.
\end{align}

Inside the gap $|\ep| < h$, we have
\begin{align}
 \int_{DE} \frac{ds}{s}\, \frac{\sqrt{\Delta_{--}}}{\sqrt{-\Delta_{+-}}}
  &= -\frac{\pi}{3} + \frac{2}{3} (2a + 1)^{3/2} K_+, \\
 \int_{DE} \frac{ds}{\lambda}\, \frac{\sqrt{\Delta_{--}}}{\sqrt{-\Delta_{+-}}}
  &= \frac{2\pi}{3} - 4 \sqrt{2a + 1} \left[ E_+ - \frac{2 + a}{3} K_+ \right], \\
 \int_H \frac{ds}{\lambda s}\, \sqrt{\Delta_{--}}
  &= 2 \pi a (1 + a).
\end{align}

Upon substitution in Eqs.\ (\ref{Xresult}) and (\ref{Psiresult}), these integrals yield the result (\ref{result}).

\subsection*{Appendix B: Integrals of Bessel functions}

The calculation of X and $\rm{\Psi}$ diagrams in real space representation boils down to the calculation of some $22$ integrals involving one Bessel function of the second kind and three Bessel functions of the first kind, Eq.\ (\ref{XYBessel}), with the arguments $p_\pm r$. Three out of four Bessel functions have the index $1$ while one of the functions has the index $0$. Since such integrals are not included in the standard reference tables, the calculation method is explained here.

Before we formulate general strategy let us consider a particular example of such an integral
\begin{equation}
\int\limits_0^\infty dr\, Y_1^+ J_0^- (J_1^-)^2=p_+ p_-^2 \int\limits_0^\infty dr\, Y_1 (p_+ r) J_0 (p_- r) J_1^2(p_- r)=p_+ p_-^2 I,
\end{equation}
where we would like to calculate the value of $I$. We note that this value is represented by the following integral
\begin{equation}
\label{example}
I=\int\limits_0^\infty dr\, Y_1 (p_+ r) J_0 (p_- r) J_1^2(p_- r)=\int\limits_0^\infty r dr\, Y_1 (p_+ r) J_0 (p_- r) \left[\frac{J_1^2(p_- r)}{r}\right].
\end{equation}
A consequence of the Gegenbauer's addition theorem for Bessel functions [see Eq.\ (16) on page 367 in Ref.\ \onlinecite{Watson}] can be applied to the expression in square brackets in Eq.\ (\ref{example}) with the result
\begin{equation}
\label{Gegenbauer}
\frac{J_1^2(p_- r)}{r}=\frac{p_-}{2\pi}\int\limits_0^\pi d\phi\,\frac{\sin^2{\phi}}{\sin{\frac{\phi}{2}}}J_1(2p_- r \sin{\tfrac{\phi}{2}}).
\end{equation}
Plugging Eq.\ (\ref{Gegenbauer}) into Eq.\ (\ref{example}) and changing the order of integrations we obtain
\begin{equation}
I=\frac{p_-}{2\pi}\int\limits_0^\pi d\phi\,\frac{\sin^2{\phi}}{\sin{\frac{\phi}{2}}} \int\limits_0^\infty r dr\,Y_1 (p_+ r) J_0 (p_- r) J_1(2p_- r \sin{\tfrac{\phi}{2}}).
\end{equation}
The integral over $r$ in the last expression is known [see Eq.\ (2.13.22.5) in Ref.\ \onlinecite{PRUD}]:
\begin{gather}
\int\limits_0^\infty r dr\,Y_1 (p_+ r) J_0 (p_- r) J_1(2p_- r \sin{\tfrac{\phi}{2}}) =-\frac{1}{2\pi p_+ p_- \sin{\tfrac{\phi}{2}}}
\begin{dcases}
1+\frac{u}{\sqrt{u^2 - 1}},
&\frac{p_- - p_+}{2p_-}>\sin{\tfrac{\phi}{2}},\phantom{\frac{p_- - p_+}{2p_-}<}\\
1,
&\frac{p_- - p_+}{2p_-}<\sin{\tfrac{\phi}{2}}<\frac{p_- + p_+}{2p_-},\\
1+\frac{u}{\sqrt{u^2 - 1}},
&\phantom{\frac{p_- - p_+}{2p_-}<}\sin{\tfrac{\phi}{2}}>\frac{p_- + p_+}{2p_-},
\end{dcases}
\\
\hspace*{-70ex}\text{where $u=\dfrac{4p_-^2\sin^2{\tfrac{\phi}{2}}+p_+^2-p_-^2}{4p_+ p_- \sin{\tfrac{\phi}{2}}}$.}
\end{gather}
Introducing the notations for the intervals $D_1=\{\phi\in[0,\pi]\mid \sin{\frac{\phi}{2}}<\frac{p_- - p_+}{2p_-}\}$ and $D_2=\{\phi\in[0,\pi]\mid \sin{\frac{\phi}{2}}>\frac{p_- + p_+}{2p_-}\}$ we write
\begin{multline}
I=-\frac{1}{4\pi^2 p_+}\left[
\int\limits_0^{\pi}d\phi\,\frac{\sin^2{\phi}}{\sin^2{\frac{\phi}{2}}}+
\int\limits_{D_{1} \bigcup D_{2}}d\phi\,\frac{\sin^2{\phi}}{\sin^2{\frac{\phi}{2}}}\frac{u}{\sqrt{u^2 - 1}}
\right]=\\
-\frac{1}{\pi^2 p_+}\left[
\int\limits_0^{\pi}d\phi\,\cos^2{\tfrac{\phi}{2}}+
\int\limits_{D_{1} \bigcup D_{2}}d\phi\,\frac{(4p_-^2\sin^2{\frac{\phi}{2}}+p_+^2-p_-^2) \cos^2{\frac{\phi}{2}}}{\sqrt{(4p_-^2\sin^2{\frac{\phi}{2}}-(p_+ + p_-)^2)(4p_-^2\sin^2{\frac{\phi}{2}}-(p_+ -p_-)^2)}}
\right],
\end{multline}
where the first integral in the last expression is readily calculated. After the substitution $t=\sin^2{(\phi/2)}$ the second integral can be expressed in terms of complete elliptic integrals of all three kinds with the moduli given by Eq.\ (\ref{moduli}) [see Eqs.\ (253.**, 257.**) in Ref.~\onlinecite{Byrd}]. With the help of known relations between elliptic integrals [see Eqs.\ (117.**) in Ref.~\onlinecite{Byrd}] one can transform the elliptic integrals of the third kind arising as the result of the calculation to the form that appeared in the left-hand side of Eq.\ (\ref{magic}). Finally all elliptic integrals of the third kind cancel out yielding
\begin{equation}
I=-\frac{1}{2\pi p_+} (1 + F_3),
\end{equation}
where $F_3$ is defined in Eq.\ (\ref{Fs}) below.

The general approach, which lets us compute all integrals involved in Eq.\ (\ref{XYBessel}), consists of the following steps:
\begin{itemize}
\item Replace the product of two Bessel functions with index $1$ and the same argument using a consequence of the Gegenbauer's addition theorem for Bessel functions [see Eq.\ (16) on page 367 in Ref.\ \onlinecite{Watson}]
\begin{equation}
\frac{J_1(\xi r)}{r}\dbinom{J_1(\xi r)}{Y_1(\xi r)}=\frac{\xi}{2\pi}\int\limits_0^\pi d\phi\,\frac{\sin^2{\phi}}{\sin{\frac{\phi}{2}}}\dbinom{J_1(2 \xi r \sin{\frac{\phi}{2}})}{Y_1(2 \xi r \sin{\frac{\phi}{2}})}.
\end{equation}
This yields the product of $r$ and three Bessel functions with different arguments.

\item Integrate over $r$ using one of these four identities ($a,b,c>0$):
\begin{gather}
\int\limits_0^\infty r dr\,J_1(a r) J_0 (b r) Y_1 (c r)=
-\frac{1}{\pi a c}
\begin{dcases}
1-\frac{u}{\sqrt{u^2 - 1}},
& b<c-a,\\
1+\frac{u}{\sqrt{u^2 - 1}},
& b<a-c,\\
1,
& |a-c|<b<a+c,\\
1+\frac{u}{\sqrt{u^2 - 1}},
& b>a+c,
\end{dcases}\\
\int\limits_0^\infty r dr\,J_1(a r) J_1 (b r) Y_0 (c r)=
\frac{1}{\pi a b}
\begin{dcases}
1-\frac{u-a/c}{\sqrt{u^2 - 1}},
& b<c-a,\\
1+\frac{u-a/c}{\sqrt{u^2 - 1}},
& b<a-c,\\
1,
& |a-c|<b<a+c,\\
1+\frac{u-a/c}{\sqrt{u^2 - 1}},
& b>a+c,
\end{dcases}\\
\int\limits_0^\infty r dr\,J_1(a r) J_1 (b r) K_0 (c r)=
\frac{1}{w_+ w_-}\times\frac{w_+ - w_-}{w_+ + w_-},\\
\int\limits_0^\infty r dr\,J_1(a r) J_0 (b r) K_1 (c r)=
-\frac{b}{c}\times\frac{1}{w_+ w_-}\left(\frac{w_+ - w_-}{w_+ + w_-}-\frac{a}{b}\right),\\
\text{with} \,\, u=\dfrac{a^2+c^2-b^2}{2ac}, \qquad
\text{and} \,\, w_{\pm}=\sqrt{\left(a\pm b\right)^2+c^2} \nonumber
\end{gather}
[see also Eq.\ (2.13.22.5) in Ref.~\onlinecite{PRUD} and Eq.\ (8.13.6) in Ref.~\onlinecite{Bateman}].
\item Integrate over $\phi$ using Ref.~\onlinecite{Byrd}. This yields complete elliptic integrals of all three kinds.
\item Reduce the moduli of complete elliptic integrals to the form Eq.\ (\ref{moduli}) and eliminate the integrals of the third kind using Eq.\ (\ref{magic}).
\end{itemize}
With the help of this approach and the use of compact notations (\ref{compact}), the complete list of relevant integrals reads (in the case $|\ep| > h$)
\begin{subequations}
\label{BesselInt}
\begin{align}
&\int\limits_0^\infty dr\, Y_1^\pm J_0^\pm (J_1^\pm)^2 = -\frac{p_\pm^2}{2 \pi}, &
&\int\limits_0^\infty dr\, Y_0^\pm (J_1^\pm)^3 = \frac{p_\pm^2}{2 \pi}, &
&\int\limits_0^\infty dr\, Y_1^- J_0^- (J_1^+)^2 = -\frac{p_+^2}{2 \pi}, \\
&\int\limits_0^\infty dr\, Y_1^+ J_0^+ (J_1^-)^2 = \frac{p_+^2}{2 \pi}-\frac{p_-^2}{\pi}, &
&\int\limits_0^\infty dr\, Y_1^- J_0^+ J_1^- J_1^+ = -\frac{p_+^2}{2 \pi}, &
&\int\limits_0^\infty dr\, Y_1^+ J_0^- J_1^- J_1^+ = -\frac{p_+^2}{2 \pi}, \\
&\int\limits_0^\infty dr\, Y_0^- J_1^- (J_1^+)^2 = \frac{p_+^2}{2 \pi}, &
&\int\limits_0^\infty dr\, Y_0^+ J_1^+ (J_1^-)^2 = \frac{p_+^2}{2 \pi}, &
&\int\limits_0^\infty dr\, Y_1^- J_0^- J_1^- J_1^+ = -\frac{p_+^2}{2 \pi}, \\
&\int\limits_0^\infty dr\, Y_1^- J_0^+ (J_1^-)^2 = -\frac{p_-^2}{2\pi}, &
&\int\limits_0^\infty dr\, Y_1^- J_0^+ (J_1^+)^2 =-\frac{p_+^2}{2\pi}(1 - F_1), &
&\int\limits_0^\infty dr\, Y_0^- J_1^+ (J_1^-)^2 = \frac{p_+^2}{2\pi}, \\
&\int\limits_0^\infty dr\, Y_0^- (J_1^+)^3 = \frac{p_+^2}{2\pi}(1 - F_2), &
&\int\limits_0^\infty dr\, Y_1^+ J_0^- (J_1^-)^2 = -\frac{p_-^2}{2\pi} (1 + F_3), &
&\int\limits_0^\infty dr\, Y_1^+ J_0^- (J_1^+)^2 =-\frac{p_+^2}{2\pi}(1 - F_2), \\
&\int\limits_0^\infty dr\, Y_1^+ J_0^+ J_1^- J_1^+ = -\frac{p_+^2}{2 \pi}(1 + F_1), &
&\int\limits_0^\infty dr\, Y_0^+ (J_1^-)^3 = \frac{p_-^2}{2\pi}(1 + F_4), &
&\int\limits_0^\infty dr\, Y_0^+ J_1^- (J_1^+)^2 = \frac{p_+^2}{2\pi}(1 - F_1).
\end{align}
In the case $|\ep| < h$, two additional integrals with modified Bessel functions are
\begin{align}
&\int\limits_0^\infty dr\, K_1 J_0^- (J_1^-)^2 = \frac{p_-^2}{4}(1 + F_5), &
&\int\limits_0^\infty dr\, K_0 (J_1^-)^3 = -\frac{p_-^2}{4}(1 - F_6).
\end{align}
\end{subequations}
The functions $F_i$, which represent combinations of complete elliptic integrals with moduli from Eq.\ (\ref{moduli}), are
\begin{subequations}
\label{Fs}
\begin{equation}
F_1=
\begin{dcases}
\frac{4 \pi a-2\sqrt{1-2 a} \left(3 E_- -(1+a) (3-a) K_-\right)}{3\pi (1-a)^2}, &1<\frac{p_-}{p_+}<3,\\
\frac{4 \pi a-2\sqrt{2 a-1} \left(3 E_-' +(2-a) a K_-'\right)}{3\pi (1-a)^2},& \frac{p_-}{p_+}>3,\\
\end{dcases}
\end{equation}
\begin{equation}
F_2=
\begin{dcases}
\frac{2\sqrt{1-2 a} \left(E_- -(1-a)^2 K_-\right)}{\pi(1-a)^2},  &1<\frac{p_-}{p_+}<3,\\
\frac{2\sqrt{2 a-1} \left(E_-' -(2-a) a K_-'\right)}{\pi(1-a)^2},& \frac{p_-}{p_+}>3,\\
\end{dcases}
\end{equation}
\begin{align}
F_3 &= \frac{4 \pi a - 4\sqrt{1+2 a}\left(3 E_+-(1-a) (3+a) K_+\right)}{3\pi(1+a)^2 },&
F_4 &= \frac{4\sqrt{1+2 a} (E_+-(1+a)^2 K_+)}{\pi(1+a)^2},\\
F_5 &= \frac{2\pi - 2\sqrt{1+2 a}\left(3 E_+-(1-a)K_+\right)}{3\pi a (1+a)},&
F_6 &= -\frac{2\sqrt{1+2 a} (E_+-(1+a) K_+)}{\pi a (1+a)}.
\end{align}
\end{subequations}
Upon substitution into Eq.\ (\ref{XYBessel}), the integrals (\ref{BesselInt}) reproduce the result (\ref{result}).

\end{document}